\documentclass[usegraphicx,useAMS,usenatbib]{mn2e}

\usepackage{amsmath}
\usepackage{amssymb}

\newcommand\Eqn[1]     {Eq.\,(\ref{#1})}

\newcommand\nn         {\nonumber}

\newcommand{\be}{\begin{equation}}
\newcommand{\ee}{\end{equation}}
\newcommand{\ba}{\begin{eqnarray}}
\newcommand{\ea}{\end{eqnarray}}

\def\FNL{{f_{\rm NL}^{\rm local}}}
\def\phiG{{\rm \Phi}_{\rm G}}
\def\phiNG{{\rm \Phi}_{\rm NG}}

\def\nhat{{\hat{\bf n}}}

\def\pp1{{\prime}}
\def\pp2{{\prime\prime}}

\def\2D{{\rm 2D}}

\def\Vu{{V_{\mu}}}

\def\pdot{\dot{\Phi}}

\def\Fka{{\mathcal F}(k)}

\def\bx{{\bf x}}

\def\bk{{\bf k}}

\def\1Loop{{\rm 1Loop}}

\def\rhob{\bar{\rho}}

\def\Mpc{\, h^{-1}{\rm Mpc}}

\def\dx{{\rm d}^3{\bf x}}

\def\del{\nabla}

\def\ga{\mathrel{\mathpalette\fun >}}
\def\fun#1#2{\lower3.6pt\vbox{\baselineskip0pt\lineskip.9pt
        \ialign{$\mathsurround=0pt#1\hfill##\hfil$\crcr#2\crcr\sim\crcr}}}

\title[ISW signature of voids and superclusters]{On the signature 
of $z\sim 0.6$ superclusters and voids in the Integrated Sachs-Wolfe effect}

\author[{\it Hern\'andez-Monteagudo \& Smith}]
       {Carlos Hern\'andez-Monteagudo$^{1}$\thanks{chm@cefca.es} \&
         Robert E. Smith$^2$\thanks{res@mpa-garching.mpg.de}\\
         {$^1$ Centro de Estudios de F\'\i sica del Cosmos de Arag\'on (CEFCA), 
           Plaza de San Juan, 1, planta 2, E-44001, Teruel, Spain}\\
         {$^2$ Max-Planck Institut f\"ur Astrophysik, 
           Karl Schwarzschild Str.1, D-85741,
           Garching bei M\"unchen, Germany}
       }
       
\begin{document}

\maketitle

\begin{abstract}
  Through a large ensemble of Gaussian realisations and a suite of
  large-volume $N$-body simulations, we show that in a standard
  $\Lambda$CDM scenario, supervoids and superclusters in the redshift
  range $z\in[0.4,0.7]$ should leave a {\em small} signature on the
  Integrated Sachs Wolfe (ISW) effect of the order $\sim 2\,\mu$K. We
  perform aperture photometry on WMAP data, centred on such
  superstructures identified from SDSS LRG data, and find amplitudes
  at the level of 8 -- 11\,$\mu$K -- thus confirming the earlier work
  of \citet{Granettetal2008b}. If we focus on apertures of the size
  $\sim3.6\degr$, then our realisations indicate that $\Lambda$CDM is
  discrepant at the level of $\sim4\,\sigma$. However, if we combine
  all aperture scales considered, ranging from 1\degr--20\degr, then the
  discrepancy becomes $\sim2\,\sigma$, and it further lowers to $\sim 0.6\,\sigma$
  if only 30 superstructures are considered in the analysis (being compatible with no ISW
  signatures at $1.3\,\sigma$ in this case).   Full-sky ISW maps generated
  from our $N$-body simulations show that this discrepancy cannot be
  alleviated by appealing to Rees-Sciama (RS) mechanisms, since their
  impact on the scales probed by our filters is negligible. We perform
  a series of tests on the WMAP data for systematics. We check for
  foreground contaminants and show that the signal does not display
  the correct dependence on the aperture size expected for a residual
  foreground tracing the density field. The signal also proves robust
  against rotation tests of the CMB maps, and seems to be spatially
  associated to the angular positions of the supervoids and
  superclusters. We explore whether the signal can be explained by the
  presence of primordial non-Gaussianities of the local type. We show
  that for models with $\FNL=\pm100$, whilst there is a change in the
  pattern of temperature anisotropies, all amplitude shifts are well
  below $<1\mu$K. If primordial non-Gaussianity were to explain the
  result, then $\FNL$ would need to be several times larger than
  currently permitted by WMAP constraints.
\end{abstract}

\date{Received XXXX; accepted YYYY}
 

%%%%%%%%%%%%%%%%%%%%%%%%%%%%%%%%%%%%%%%%%%%%%%%%%%%%%%%

\voffset -1.3cm

\section{Introduction}

Over the last fifteen years, evidence has been mounting from various
cosmological probes to support the case for an accelerating universe.
It can be argued that the first compelling evidence for this arose
from the study of the light curves of distant type Ia supernovae
\citep{Riessetal1998short,Perlmutteretal1999short}. Currently, the
strongest support for this picture comes from the combination of
observations of the Cosmic Microwave Background radiation (hereafter
CMB), and from measurements of the clustering of galaxies. From the
CMB side, the Wilkinson Microwave Anisotropy Probe (hereafter, WMAP)
experiment
\citep{Spergeletal2003short,Spergeletal2007short,Komatsuetal2011short}\footnote{\tt
  http://map.gsfc.nasa.gov} provided a precise measurement of the
angular size of the sound horizon at recombination, which supported
the case for a spatially flat universe. On the clustering side, data
from surveys like the 2-degree Field Redshift Survey
\citep{Coleetal2005short} and the Sloan Digital Sky Survey (hereafter
SDSS) \citep{Tegmarketal2006short} required the density in matter to
be sub-critical, hence leading to the inference that $\sim$70 per cent of the
current energy density of the Universe is in a form of energy that
behaves like a cosmological constant, and so acts as a repulsive
gravitational force. The energy density driving the accelerated
expansion is unknown and so has been dubbed {\it Dark Energy}
(hereafter DE). Uncovering the true physical nature of the DE is one
of the main targets for many ongoing and upcoming surveys of the
Universe.

Standard linear cosmological theory states that if the Universe
undergoes a late-time phase of accelerated expansion, then
gravitational potential wells on very large scales ($\ga
100$\,h$^{-1}$Mpc) will decay. This evolution of the potential wells
introduces a gravitational blueshift in the photons of the CMB that is
known as the Integrated Sachs-Wolfe effect (ISW). This effect
constitutes an alternative window to DE, and can be directly measured
by cross-correlating CMB maps with a set of tracers for the density
field, which sources the potentials \citep{CrittendenTurok1996}. As
soon as the first data sets from WMAP were released, several works
claimed detections of the ISW at various levels of significance
\citep[][]{Scrantonetal2003short,Fosalbaetal2003,BoughnCrittenden2004,
  Fosalbaetal2004,Noltaetal2004,Afshordietal2004,Padmanabhanetal2005c,
  Cabreetal2006,Giannantonioetal2006,vielvaetal2006,McEwenetal2007}.
Subsequent analysis has led some researchers to be more cautious about
interpreting these early detections
\citep{HernandezMonteagudoetal2006,Rassatetal2007,Bielbyetal2010,
  LopezCorredoiraetal2010,HernandezMonteagudo2010,PeacockFrancis2010}. As
emphasized by \citet{HernandezMonteagudo2008}, the true ISW effect
should only be detectable for deep galaxy surveys that cover a
substantial fraction of the sky. However an erroneous interpretation
of ISW cross correlation studies may be obtained from systematic
errors, such as residual point source emission in CMB maps or presence
of spurious galaxy auto-power on large angular scales \citep[see also][for details
  on the WMAP -- NVSS cross-correlation
  analysis.]{HernandezMonteagudo2010}. In particular, the issue of
excess power on large scales has been noted in several works
\citep{Hoetal2008,HernandezMonteagudo2010,Thomasetal2011,
  Giannantonioetal2012}, but it is not yet fully accounted for.

Amongst the subsequent ISW cross-correlation studies in the
literature, there is the particularly puzzling work of
\citet[][hereafter G08]{Granettetal2008b}. Their analysis yielded one
of the highest detection significances in the literature. G08
implemented the following novel approach: they produced a catalogue of
superclusters and supervoids from SDSS data, and stacked WMAP-filtered
data on the positions of these structures with apertures of the order
$\sim4\degr$. They obtained a $\sim4\,\sigma$ ISW detection. Unlike
previous works on the subject, the analysis focused on a particular
subset of the available large-scale structure (hereafter LSS)
data. Subsequent studies have investigated the origin of the signal
and assessed its compatibility with the $\Lambda$CDM scenario,
\citep{PapaiSzapudi2010,Papaietal2011,Nadathuretal2012}. While some
works found the G08 results compatible, others found the measured
amplitudes too high to be consistent
\citep{Granettetal2009,Nadathuretal2012,Flenderetal2012,Inoueetal2010,Inoue2012}. Currently, the G08 results
remain unexplained.

In this work we shall attempt to shed new light on this problem.  The
paper breaks down as follows: In \S\ref{sec:theory} we give a brief
theoretical overview of the ISW effect, underlining the expectation
for the ISW signal from a cross-correlation analysis of the data used
in G08. In \S\ref{sec:Granett} we perform a cross-check of the G08
results.  In \S\ref{sec:tests} we test whether the G08 results are
consistent with expectations for the $\Lambda$CDM model. This is
achieved by using a large ensemble of Gaussian Monte-Carlo
realisations. We also generate full-sky nonlinear ISW maps by
ray-tracing through a suite of $N$-body simulations. In
\S\ref{sec:discussion} we determine the level of significance at which
the results disagree with the $\Lambda$CDM paradigm. We explore
systematic errors in the foreground subtraction for WMAP. We also
investigate whether the excess signal is consistent with primordial
non-Gaussianities of the local type. Finally, in
\S\ref{sec:conclusions} we summarize our findings and conclude.
While this paper was in the process of submission, a parallel work on this subject from 
\citet{Flenderetal2012} appeared in the internet. This work is reaching to similar conclusions 
to ours in some of the issues addressed in this work.

Unless stated otherwise, we employ a reference cosmological model
consistent with WMAP7 \citep{Komatsuetal2011short}: the energy-density
parameters for baryons, CDM and cosmological constant are $\Omega_{\rm
  b}=0.0456$, $\Omega_{\rm cdm}=0.227$, $\Omega_{\Lambda}=0.7274$; the
reduced Hubble rate is $h=0.704$; the scalar spectral index is $n_{\rm
  S}=0.963$; the rms of relative matter fluctuations in spheres of
$8\,h^{-1}$\,Mpc radius is $\sigma_8=0.809$, and the optical depth to
last scattering is $\tau=0.087$.

%%%%%%%%%%%%%%%%%%%%%%%%%%%%%%%%%%%%%%%%%%%%%%%%%%%%%%%

\section{Theoretical perspective}\label{sec:theory}

%%%%%%%%%%%%%%%%%%%%%%%%%%%%%%%%%%%%%%%%%%%%%%%%%%%%%%%

\subsection{The ISW effect}

Observed CMB photons are imprinted with two sets of fluctuations:
primary anisotropies, sourced by fluctuations at the last-scattering
surface, and secondary anisotropies, induced as the photon propagates
through the late-time clumpy Universe. The physics of the primary
anisotropies is well understood \citep[][]{Dodelson2003,Weinberg2008}.
There are a number of physical mechanisms that give rise to the
generation of secondary anisotropies \citep[for a review
  see][]{PlanckBlueBook} and one of these is the redshifting of the
photons as they pass through evolving gravitational potentials. The
linear version of this effect is termed the Integrated Sachs-Wolfe
effect \citep{SachsWolfe1967} and its nonlinear counter-part is termed
the Rees-Sciama effect \citep{ReesSciama1968}.

The observed temperature fluctuation induced by gravitational
redshifting may be written as \citep{SachsWolfe1967}:
\be \frac{\Delta T(\nhat)}{T_0} = -{2 \over c^2}\int_{t_{\rm
ls}}^{t_{0}}dt \,\dot{\Phi}(\nhat,\chi) \ \label{eq:ISW} ,\ee
where $\nhat$ is a unit direction vector on the sphere, $\Phi$ is the
dimensionless metric perturbation in the Newtonian gauge, which
reduces to the usual gravitational potential on small scales, the
`over dot' denotes a partial derivative with respect to the coordinate
time $t$ from the FLRW metric, $\chi$ is the comoving radial geodesic
distance $\chi=\int cdt/a(t)$, and so may equivalently parameterize
time. The symbols $t_0$ and $t_{\rm ls}$ denote the time at which the
photons are received and emitted, i.e. the present time and last
scattering. $c$ is the speed of light and $a(t)$ is the dimensionless
scale factor.

On scales smaller than the horizon, relevant to our simulation boxes,
the perturbed Einstein equations in Newtonian gauge lead to a
perturbed Poisson equation. This enables us to relate potential and
matter fluctuations \citep{Dodelson2003}:
\be \nabla^2\Phi(\bx;t)=4\pi G\rhob(t)\delta(\bx;t) a^2\!(t) \ , \ee
where $\rhob(t)$ is the mean matter density in the Universe and the
density fluctuation is defined $\delta(\bx;t)\equiv
[\rho(\bx,t)-\rhob(t)]/\rhob(t)$. This equation may most easily be
solved in Fourier space:
\be \Phi(\bk;t) = -4\pi G\rhob(t)a^2(t)\frac{\delta(\bk;t)}{k^2}\ . \ee
Differentiation of the above expression gives 
\be \dot{\Phi}(\bk;t) =
\frac{3}{2}\Omega_{\rm m0} H_0^2 k^{-2} 
\left[ 
\frac{H(t)}{a(t)}\delta(\bk;t)-\frac{\dot{\delta}(\bk;t)}{a(t)}
\right] \label{eq:ISW1} \ ,
\ee
where we used the fact that $[a^3(t)\rhob(t)]$ is a time-independent
quantity in the matter-dominated epoch and \mbox{$\Omega_{\rm m}\equiv
\Omega_{\rm cdm}+\Omega_{\rm b}$}. In the above, we also defined
$H(t)\equiv \dot{a}(t)/a(t)$ and \mbox{$\Omega_{\rm m}(t)\equiv
  \rhob(t)/\rho_{\rm crit}(t)$}, with $\rho_{\rm crit}(t)=3H^2(t)/8\pi
G$. All quantities with a subscript $0$ are to be evaluated at the
present epoch. In the linear regime, density perturbations scale as 
$\delta(\bk,a)=D(a)\delta(\bk,a_0)$. Inserting this relation into
\Eqn{eq:ISW1} gives,
\be \dot{\Phi}(\bk;t) 
= 
\frac{3}{2}\Omega_{\rm m0} H_0^2 k^{-2} 
\frac{H(t)}{a(t)}  \left[ 1 -\frac{d\log D}{d\log a}\right] \delta(\bk;a) \label{eq:ISW1b} \ .
\ee
In the matter-dominated phase, the Universe expands as in the
Einstein-de Sitter case and consequently density perturbations scale
as $D(a)\propto a$. Thus, for most of the evolution of the late-time
Universe the bracket, $\left[1 -d\log D/d\log a\right]$ is close to
zero. In the $\Lambda$CDM model it is only at relatively late times
that this term is non-zero.

%%%%%%%%%%%%%%%%%%%%%%%%%%%%%%%%%%%%%%%%%%%%%%%%%%%%%%%

\begin{figure*}
\centerline{
\includegraphics[width=17cm]{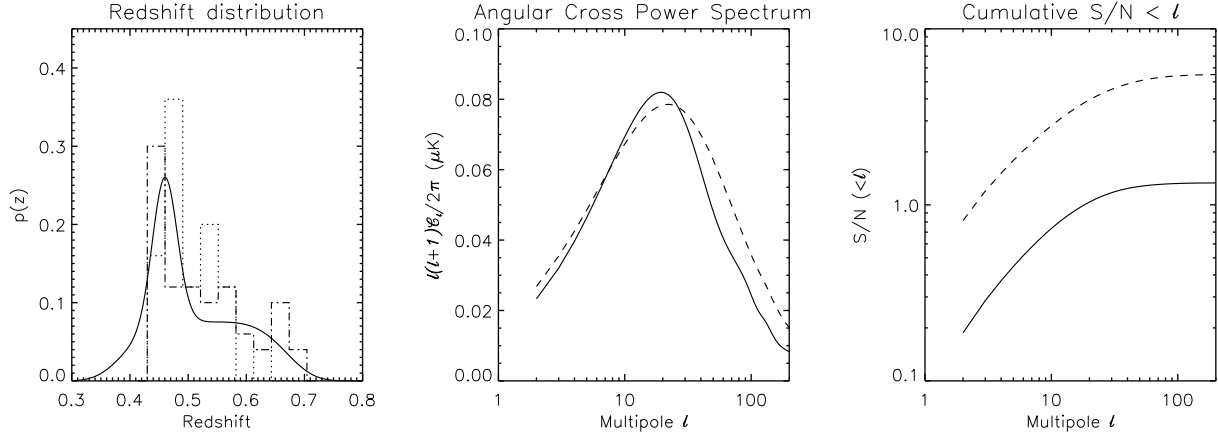}}
\caption{\small{ {\em Left panel}: Redshift distribution of supervoids
    and superclusters for the catalogue provided by G08 from the SDSS
    DR6 data. The dotted histogram denotes supervoids and the
    dot-dashed histogram denotes superclusters.  The solid line
    represents an analytic approximation, which is a compromise
    between both histograms. {\em Middle panel}: angular cross-power
    spectrum between ISW and the projected density. The solid line
    denotes the predictions for both supervoids and superclusters,
    modulo the bias being unity. The dashed curve represents what one
    would expect for QSO's in the NVSS survey. {\em Right panel}:
    Cumulative ${\mathcal S}/{\mathcal N}$ below a given multipole
    $\ell$. The solid line denotes the prediction for matter within
    the redshift distribution sampled by the G08 catalogue. The dashed
    line presents the predictions for QSO's in NVSS. }
\label{fig:s2n}}
\end{figure*}

%%%%%%%%%%%%%%%%%%%%%%%%%%%%%%%%%%%%%%%%%%%%%%%%%%%%%%%

Alternatively, in the nonlinear regime $\delta(k,a)\ne
D(a)\delta(k,a_0)$, and this gives rise to additional sources for the
heating and cooling of photons \citep{Smithetal2009,Caietal2009,Caietal2010}.

%%%%%%%%%%%%%%%%%%%%%%%%%%%%%%%%%%%%%%%%%%%%%%%%%%%%%%%

\subsection{Expectations for cross-correlation analysis}

The expected signal for the ISW-density tracer cross-correlation
analysis has been described by several authors
\citep{CrittendenTurok1996,HernandezMonteagudo2008,Smithetal2009} and
here we simply quote the main results.  For a survey of galaxies (or
more generically, objects that are biased density tracers) the
ISW-density angular cross-power spectrum can be written:
\ba
C_{\ell}^{\rm g, ISW} \hspace{-0.2cm} & = &\!\! 
\left( \frac{2}{\pi}\right)\int_0^{\infty} \;k^2dk\;  P_m(k)
  \; \nn \\
\!\!& \times &\!\!\! \int_{0}^{\infty} d\chi_1\; j_{\ell} (k\chi_1) \chi_1^2
W_{\rm g}(\chi_1)n_{\rm g}(\chi_1)b(\chi_1,k)D(\chi_1) \nn \\
\!\!& \times & \!\!\! \int_{0}^{\infty} d\chi_2\; j_{\ell} (k\chi_2) 
\frac{-3\Omega_{\rm m}H_0^2}{k^2} \;\frac{d(D(\chi_2)/a(\chi_2))}{d\chi_2} \ ,
\label{eq:x_iswrho1}
\ea
where $P_m(k)$ denotes the linear matter power spectrum at the present
time, the super-script ${\rm g}$ refers to the objects ({\it galaxies}
in most cases, but in ours it will refer to voids or superclusters)
probing the gravitational potential wells. In the above $n_{\rm
  g}(\chi)$ denotes the average comoving object number density;
$W_{\rm g}(\chi)$ denotes the instrumental window function providing
the sensitivity of the instrument to the objects at distance $\chi$;
the $j_{\ell}(x)$ are the usual spherical Bessel functions of order
$l$; and $b(\chi,k)$ denotes the {\em bias} of the tracer population
with respect to the matter density field.  This last factor may be
both a function of time $\chi$ and scale $k$ \citep{Smithetal2007},
but for simplicity we shall simply assume that the bias equals unity
for the objects in our catalogue.

In this measurement the primary CMB signal acts as {\em noise}. If we
take this into account and also the variance due to the density field,
then the ${\mathcal S}/{\mathcal N}$ with which we expect to detect the
cross-correlation for a given multipole is \citep[e.g.,][]{CrittendenTurok1996}:
\ba
\left(\frac{\mathcal S}{\mathcal N} \right)^2_{\ell} & = &
\frac{\left(C_{\ell}^{\rm g,ISW}\right)^2\times 
(2\ell+1)f_{\rm sky}}{C_{\ell}^{\rm CMB}(C_{\ell}^{\rm g} +
   1/{\bar n}_{\rm g})+\left(C_{\ell}^{\rm g,ISW}\right)^2} \ ,
\label{eq:s2n1} 
\ea
where $C_{\ell}^{\rm CMB}$ and $C_{\ell}^{\rm g}$ represent the auto
power spectra of the temperature fluctuations and the density tracers;
$\bar{n}_{\rm g}$ denotes the average {\em angular} object number
density and hence the term $1/{\bar n}_{\rm g}$ denotes the
shot-noise. In our forecast we shall neglect this term (hence the
predicted significance will be slightly over optimistic). The factor
$f_{sky}$ refers to the fraction of the sky jointly covered by both
the CMB and the object surveys. The cumulative ${\mathcal S}/{\mathcal
  N}$ for all harmonics smaller than $l$ can then be written:
\be
\frac{\mathcal S}{\mathcal N} (<\ell) = \sqrt{\sum_{\ell'=2}^{\ell} \left( \frac{\mathcal S}{\mathcal N}\right)^2_{\ell'}},
\label{eq:s2n2}
\ee
where the monopole and dipole have not been included.

%%%%%%%%%%%%%%%%%%%%%%%%%%%%%%%%%%%%%%%%%%%%%%%%%%%%%%%

\subsection{Expectations for ISW from data used in Granett et al.}

We now turn to the question of what ${\mathcal S}/{\mathcal N}$ should
G08 have expected to find in the their data. G08 used a sample of 1.1
million Luminous Red Galaxies (hereafter LRG) selected from the SDSS
data release 6 (hereafter DR6)
\citep{AdelmanMccarthyetal2008short}. This sample covered roughly 7500
square degrees and spanned a redshift range of $(0.4<z<0.75)$. From
this sample they identified regions as supervoids or superclusters
using the algorithms {\tt ZOBOV} and {\tt VOBOZ}
\footnote{\tt http://skysrv.pha.jhu.edu/$\sim$neyrinck/voboz/},
respectively, \citep{Neyrinck2008, Neyrincketal2005}. 
The significance of these regions was chosen to be at
least at the $2\sigma$-level relative to a Poisson sample of
points. G08 selected the largest 50 superclusters and supervoids for
their analysis. Their catalogue is publically available\footnote{\tt
  http://ifa.hawaii.edu/cosmowave/supervoids/}.

In the left panel of Fig.~\ref{fig:s2n}, the dot-dashed and dotted
histograms denote the redshift distributions of supervoids and
superclusters, respectively.  The thick solid line represents an
analytic fit that we have constructed, which attempts to be a
compromise between the two. Our fitting function extends to slightly
lower redshifts than the G08 catalogue, and this will translate into a
slightly higher ISW prediction.

The middle panel of Fig.~\ref{fig:s2n} presents the angular
cross-power spectrum as predicted by linear theory for the
void/supercluster catalogue (solid line). Recall that we are taking
$b=1$ for both voids and clusters. In reality the void/cluster regions
will be anti-biased/biased and so the two signals will
differ. However, since here we are more concerned with the ${\mathcal
  S/N}$ this simplification does not matter, especially since we are
neglecting the effects of shot-noise on the cross-spectra. We also
point out that our predictions for the ISW-induced cross correlation
signal of superclusters and supervoids in the SDSS sample are quite
similar to the predictions for the cross-correlation of AGNs ($z<2$)
in the NVSS catalogue \citep[after adopting the model of][see the
  dashed line in the middle panel of
  Fig.~\ref{fig:s2n}]{Hoetal2008}.

The right panel of Fig.~\ref{fig:s2n} presents the predictions for
the cumulative ${\mathcal S}/{\mathcal N}$ for the SDSS supercluster
and supervoid analysis. These predictions (solid black line) show
that, for a $\Lambda$CDM model, one should expect no more than $\simeq
1.3\sigma$ significance. In contrast, the prediction for NVSS (dashed
line) is close to $\sim$5.5 (obtained after also neglecting shot
noise).  This is not surprising, since the G08 supervoid and
supercluster catalogue is relatively shallow, spanning the redshift
range $(0.4<z<0.7)$, and covers only a modest fraction of the sky
$(f_{\rm sky}\simeq 0.18)$.  The NVSS is instead significantly deeper
and wider. We hence conclude that had G08 applied a standard ISW
cross-correlation analysis to their data, then in the framework of the
$\Lambda$CDM model, there would have been very little chance for
detecting any genuine signal at high significance.

%%%%%%%%%%%%%%%%%%%%%%%%%%%%%%%%%%%%%%%%%%%%%%%%%%%%%%%

\section{Cross-checking Granett et al.}\label{sec:Granett}

%%%%%%%%%%%%%%%%%%%%%%%%%%%%%%%%%%%%%%%%%%%%%%%%%%%%%%%

\subsection{CMB data}

The WMAP experiment scanned the CMB sky from 2001 until 2010 in five
different frequencies, ranging from 23 GHz up to 94 GHz. The angular
resolution in each band improves with the frequency, but it remains
better than one degree in all bands. The ${\mathcal S}/{\mathcal N}$
is greater than one for multipoles $\ell<919$
\citep{Jarosiketal2010short}, and in particular, on the large scales
of interest for ISW studies, the galactic and extragalactic foreground
residuals are below the 15$\mu$K level outside the masked regions
\citep{Goldetal2011short}.

We concentrate our analysis on the foreground-cleaned maps
corresponding to bands Q (41GHz), V (61GHz) and W (94GHz), after
applying the conservative foreground mask KQ75y7, which excludes
$\sim$25\% of the sky. At the scales of interest, instrumental noise
lies well below cosmic variance and foreground residuals, and hence
will not be considered any further. The ISW is a thermal signal whose
signature should not depend upon frequency and hence should remain
constant in the three frequency channels. All of the WMAP data
employed in this analysis were downloaded from the {\tt LAMBDA}
site\footnote{\tt http://lambda.gsfc.nasa.gov}.
  
%%%%%%%%%%%%%%%%%%%%%%%%%%%%%%%%%%%%%%%%%%%%%%%%%%%%%%%

\subsection{Supercluster and Supervoid data}

For our tracers of the LSS, we use the same supercluster and supervoid
catalogue as used by G08. As described earlier, the catalogue was
constructed after applying the {\tt ZOBOV} and {\tt VOBOZ} algorithms
to search for supervoids and superclusters in the LRG sample extracted
from SDSS DR6, respectively. G08 used the 50 largest supervoids and
superclusters. They claimed that this cut yielded the highest
statistical significance, in that it minimized the contamination from
spurious objects, whilst at the same time it provided sufficient
sampling to beat down the intrinsic CMB noise.

%%%%%%%%%%%%%%%%%%%%%%%%%%%%%%%%%%%%%%%%%%%%%%%%%%%%%%%

\subsection{Methodology}\label{sec:method}

In their approach G08 have applied a top-hat compensated filter or
Aperture Photometry (AP) method to the CMB map(s) positions of voids
and superclusters. This filter subtracts the average temperature
inside a ring from the average temperature within the circle limited
by the inner radius of the ring. In order to have equal areas in both
cases, the choice of the outer radius of the ring is $\sqrt{2}\, R$,
with $R$ the inner radius of the ring. In this way, fluctuations of
typical size $R$ are enhanced against fluctuations at scales smaller
or larger than such radius. Although G08 present results for apertures
ranging from 3\degr up to 5\degr, most of the conclusions are driven
from the $R=4\degr$ choice, for which highest statistical significance
is achieved: they find that AP stacks on the position of voids
(superclusters) yield a decrement (increment) of $\sim -11.3$\,$\mu$K
($7.9$\,$\mu$K) at 3.7 (2.6)\,$\sigma$ significance level. However, it
turns out that, according to G08, the typical size of clusters and
voids are $\sim$ 0.5\degr and 2\degr, respectively, which seem to lie
at odds with the aperture choice of 4\degr. Potentials are known to
extend to larger scales than densities, and it is {\em a priori}
unclear which aperture radius should be used.  This fact motivates a
systematic study in a relatively wide range of aperture radii.

%%%%%%%%%%%%%%%%%%%%%%%%%%%%%%%%%%%%%%%%%%%%%%%%%%%%%%%
 
\begin{figure}
\includegraphics[width=8.5cm]{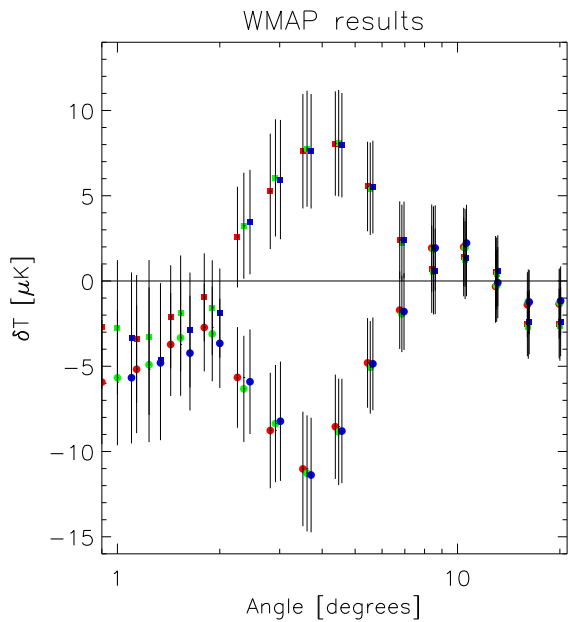}
  \caption{AP filter outputs versus aperture size. Red, green and
    blue symbols refer to WMAP's Q, V and W bands,
    respectively. Squares and circles refer to superclusters and
    voids, respectively.}
\label{fig:wmapdata}
\end{figure}

%%%%%%%%%%%%%%%%%%%%%%%%%%%%%%%%%%%%%%%%%%%%%%%%%%%%%%%

\subsection{Stacking analysis on real data}

We apply the G08 method on the Q, V, and W bands of the WMAP data
using the SDSS supercluster and supervoid catalogues. We have
considered AP filters in 15 logarithmically spaced bins in the angular
range 1\degr--20\degr. The filters were placed on the centers of the
objects, as they are provided by the catalogue.

Figure~\ref{fig:wmapdata} displays the results for the stacked signal
as a function of the AP filter aperture size in degrees. The red,
green and blue symbols refer to results from the Q, V and W bands,
respectively. We clearly see that there is practically no frequency
dependence. The error bars are computed after repeating the analysis
on 30 random sets of 50 objects placed in the un-masked region of the
sky. Our findings are in good agreement with those of G08: voids and
supercluster regions yield a slightly asymmetric pattern, with voids
rendering amplitudes of $\simeq -11$\,$\mu$K for apertures of $\simeq
3.6\degr$, and superclusters giving rise to increments of $\simeq
9$\,$\mu$K at that same scale. In these two cases, the significance is
about -3.3\,$\sigma$ and 2.3\,$\sigma$, which on combination yields a
combined significance $\sim4\,\sigma$. Again, this is in good
agreement with G08. Furthermore, the scale showing highest ${\mathcal
  S}/{\mathcal N}$ is $\lesssim4\degr$, as the significance rapidly
drops for smaller and larger apertures. Intuitively, this seems to be
in contradiction with the idea of ISW fluctuations being large-scale
anisotropies, since in such case one would expect to attain high
${\mathcal S}/{\mathcal N}$ also for moderately large ($\gtrsim
5\degr$--$10\degr$) apertures.

%%%%%%%%%%%%%%%%%%%%%%%%%%%%%%%%%%%%%%%%%%%%%%%%%%%%%%%

\begin{figure}
\includegraphics[width=8.5cm]{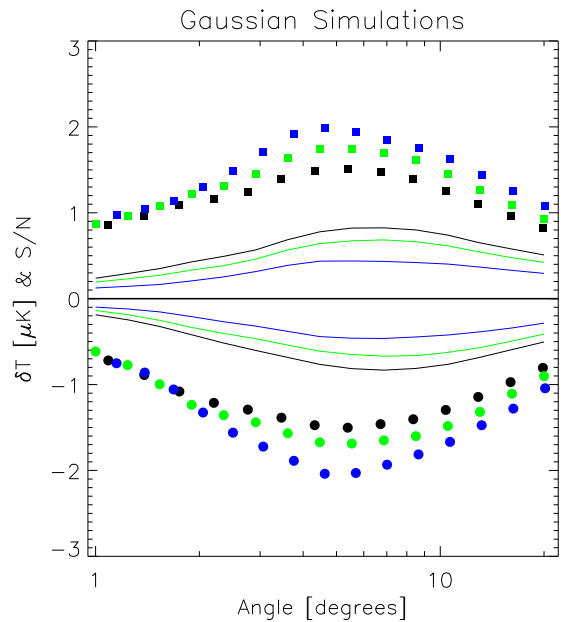}
  \caption{AP filter outputs for the Gaussian realisations of the V
    band of WMAP. Results for realisations of other bands are virtually
    identical to these. Black, green and blue symbols and curves refer
    to density threshold choices of $\nu=$ 2.5, 2.8 and 3.2,
    respectively. Circles correspond to voids and squares to
    superclusters, while solid lines display the
    $\mathcal{S}/\mathcal{N}$ at each scale.}
\label{fig:gsims}
\end{figure}

%%%%%%%%%%%%%%%%%%%%%%%%%%%%%%%%%%%%%%%%%%%%%%%%%%%%%%%

\begin{figure*}
\centering{
  \includegraphics[width=8cm,clip=,angle=90]{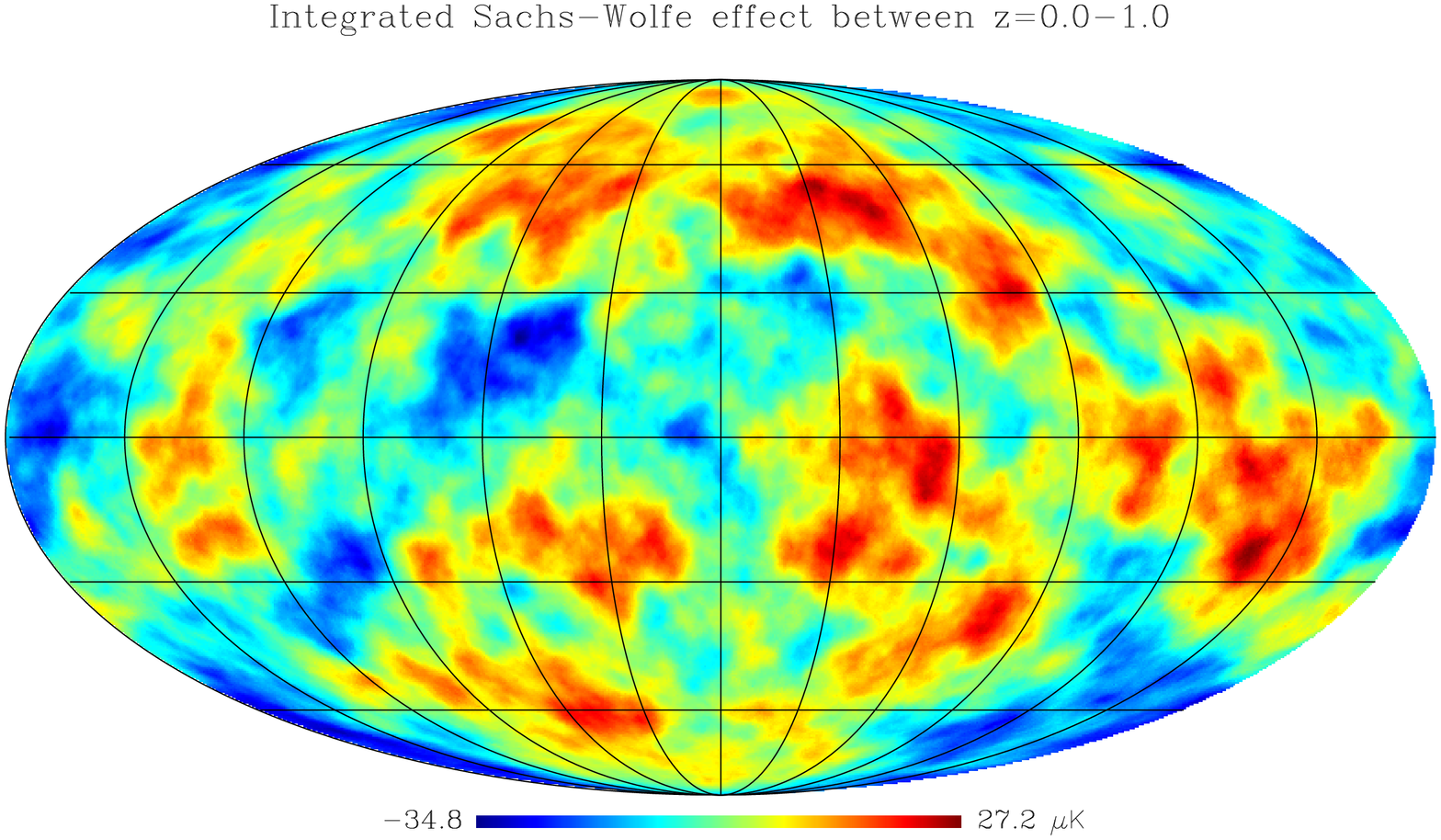}}
\centering{
  \includegraphics[width=8cm,clip=,angle=90]{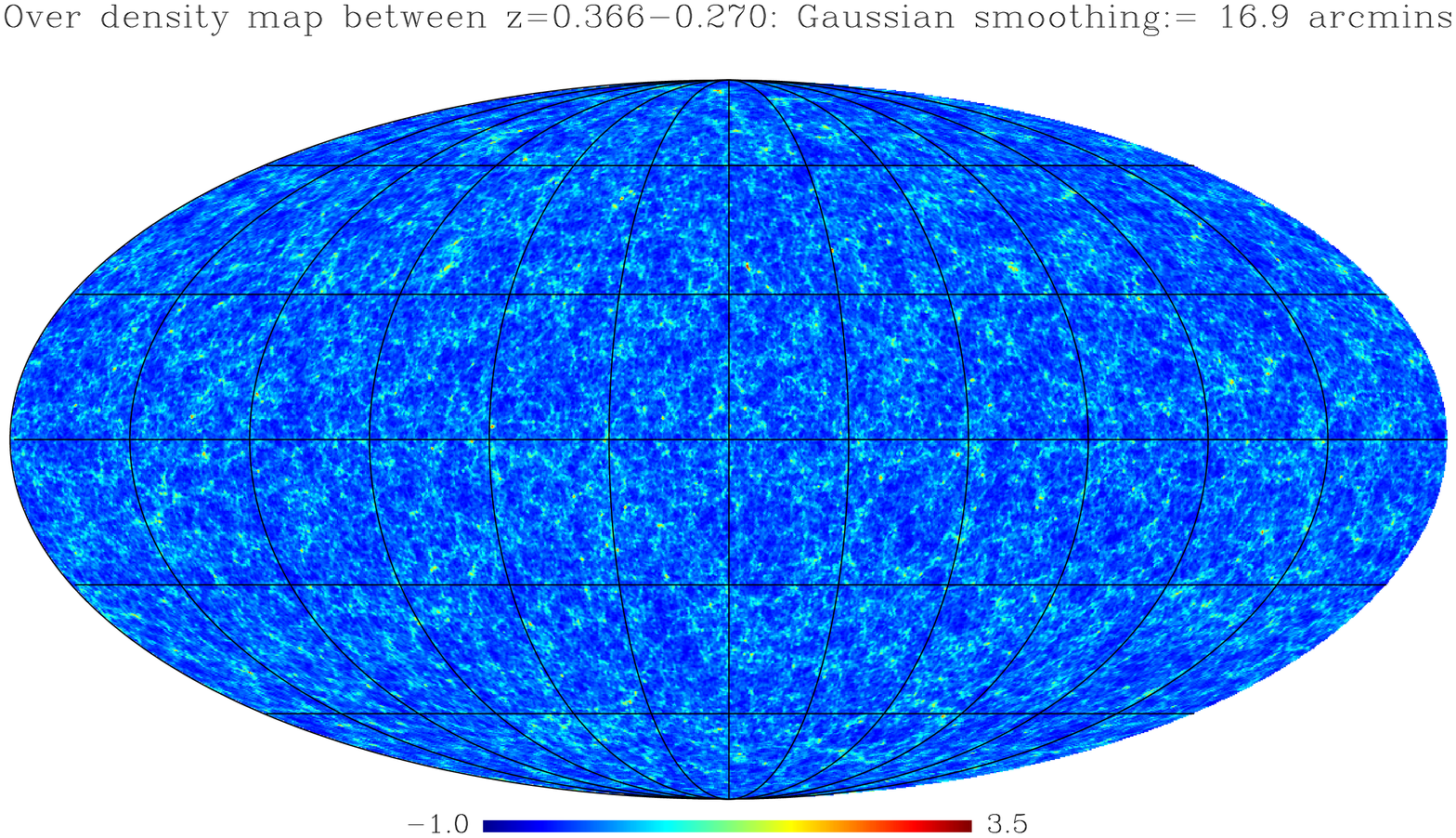}}
\caption{\small{{\em Top panel}: Full-sky, nonlinear ISW map arising
    from structures between $z=0.0$ and $z=1.0$, estimated from one of
    the {\tt zHORIZON} simulations. The sky map was pixelated using
    the {\tt HEALPix}
    package with a resolution of $N_{\rm side}=256$,
    which corresponds to 786,432 pixels. {\em Bottom panel:} full-sky
    projected over-density map for the distribution of cold dark
    matter structures located in the redshift shell $z=[0.260,0.366]$
    for the same $N$-body simulation. The map was pixelated at the
    same resolution as the ISW map and smoothed with a Gaussian filter
    of scale ${\rm FWHM}=16.9$ arcmins (roughly the angular size
    subtended by $1\Mpc$ at $z=0.3$).  In both panels the graticule
    scale indicates 30\degr divisions.}}
\label{fig:ISWMap}
\end{figure*}

%%%%%%%%%%%%%%%%%%%%%%%%%%%%%%%%%%%%%%%%%%%%%%%%%%%%%%%

\section{Are the results consistent with the vanilla $\Lambda$CDM universe?}\label{sec:tests}

%%%%%%%%%%%%%%%%%%%%%%%%%%%%%%%%%%%%%%%%%%%%%%%%%%%%%%%

\subsection{Gaussian realisations}\label{ssec:gaussian}

What is not clear from the analysis of the previous section, is
whether the $\sim 4\sigma$ detection from the stacked
supercluster/supervoid regions is consistent with what one expects
from the standard $\Lambda$CDM model \citep[][]{Komatsuetal2011short}.
We now attempt to understand the dependence of the expected signal and
its errors on the filter size. To that end, we repeat the above
analysis on a set of Gaussian realisations of both the LSS distribution
and the corresponding CMB temperature anisotropy distribution which
would result in our $\Lambda$CDM model.

The CMB maps are constructed in a two-step process: first, we generate
a Gaussian map of projected density, following the angular power
spectrum built upon the redshift window function $W_{\rm g}(r)$
displayed in the left panel of Fig.~\ref{fig:s2n}. This density map
is used for {\it (i)} constructing a supercluster and supervoid
catalogue (see below), and {\it (ii)} generating an ISW
component. This ISW component is correlated to the density map as
predicted by Eq.~\ref{eq:x_iswrho1}, \citep[see, e.g.,][for
  details.]{Cabreetal2006,HernandezMonteagudo2008} Second, we generate
the primary anisotropy signal at $z\simeq 1\,050$. This is taken to be
completely uncorrelated with respect to the projected density
map. This is not exactly true due to the lensing of the CMB, but it is
still a very good approximation on the large angular scales of
interest in this study.  The simulated ISW map is then directly
co-added to the primary CMB temperature map. We have checked that the
cross-correlations of the simulated density and CMB maps are in direct
agreement with a numerical evaluation of Eq.~\ref{eq:x_iswrho1}.

All of the simulated sky maps were generated using the equal area
pixelization strategy provided by {\tt HEALPix}\footnote{URL site: {\tt http://healpix.jpg.nasa.gov}}, \citep{Gorskietal2005}. 
 We take the pixel scale to be $\simeq$
15\arcmin, corresponding to a {\tt HEALPix} resolution parameter of
$N_{\rm side}=256$.  For each simulated LSS map we smooth the map
with a Gaussian aperture of FWHM$\simeq$2\degr, and identify those
peaks and troughs which exceed a given threshold $\nu|\sigma|$ as
being associated with supervoids and superclusters, where $\sigma$ in
this context refers to the density field rms. The threshold $\nu$ is
chosen to have an object density similar to that of superclusters and
voids in the real catalogue under the SDSS footprint.  Note that this Gaussian smoothing takes place only at the step of `identifying' superstructures in the density maps. To check the
dependence on the threshold choice, we bracket the preferred value of
$\nu$ with two other values, one above and one below it. The final
choice for the $\nu$ value set was 2.5, 2.8 and 3.2.

To each simulated CMB sky, we also add a noise realization following
the anisotropic noise model provided in the {\tt LAMBDA} site. We then
exclude pixels in accordance with the intersection of the WMAP KQ75y7
sky mask and the SDSS DR6 data footprint.

Figure~\ref{fig:gsims} presents the ensemble-averaged results obtained
from 5000 Gaussian Monte-Carlo realisations. As expected, for the three
adopted thresholds, overdensities (or positive excursions in the
projected density map) yield a positive signal for the stacked
aperture analysis (as displayed by the squares in the plots), whereas
underdensities yield negative ones (circles in the plot). The
Monte-Carlo realisations also enable us to compute the variance on the
measurements. On taking the ratio of the mean signal and the rms
noise, we obtain direct estimates of the ${\mathcal S}/{\mathcal N}$
for each given aperture bin. The coloured solid lines in the plot
present our direct measurements of the ${\mathcal S}/{\mathcal
  N}$. Thus we clearly see that the scatter induced by the CMB
generated at $z\simeq 1050$ is the dominant source of noise, keeping
the ${\mathcal S}/{\mathcal N}$ for each angular bin below unity.

For lower thresholds there is more area covered and intuitively one
would expect a higher ${\mathcal S}/{\mathcal N}$, as it seems to be
the case. Our realisations also provide higher ISW amplitudes for
higher thresholds, and this makes sense since deeper voids/potential
wells should have a stronger impact on CMB photons. Nevertheless, in
all cases typical amplitudes remain at the level of
$1$--$2\,\mu$K. The aperture at which the AP outputs provide the
highest amplitude does not show any strong dependence upon the
threshold $\nu$, and seems to lie in a wide angle range within
[3\degr, 8\degr]. For apertures larger than 10\degr, the ${\mathcal
  S}/{\mathcal N}$ for the lowest threshold starts dropping slowly,
and becomes half of its maximum value at an aperture of 20\degr. For
higher thresholds this decrease is found to be even shallower.

%%%%%%%%%%%%%%%%%%%%%%%%%%%%%%%%%%%%%%%%%%%%%%%%%%%%%%%

\subsection{Generation of nonlinear ISW and density maps from 
$N$-body simulations}\label{sec:zhor_analysis}

%%%%%%%%%%%%%%%%%%%%%%%%%%%%%%%%%%%%%%%%%%%%%%%%%%%%%%%

The previous subsection has shown that the Gaussian realizations of
the $\Lambda$CDM universe are in tension with the excess signal found
by G08. One weak point in the above analysis is that the density and
late-time potential field are not necessarily well described by a
Gaussian process, since nonlinear evolution under gravity drives the
initially Gaussian distribution of density fluctuations towards one
that is non-Gaussian at late times. In order to test whether nonlinear
evolution could explain the excess signal seen by G08, we now turn to
the challenge of constructing fully nonlinear maps of the density
field and ISW effect from $N$-body simulations.

The 8 simulations that we employ for this task are a sub-set of the
{\tt zHORIZON} simulations. These simulations were used in
\citet{Smithetal2009} to calculate the expected ISW-cluster
cross-power spectra. In brief, each simulation follows the
gravitational evolution of $N=750^3$ dark matter particles in a box of
comoving size $L=1500\Mpc$.  The cosmological model employed was a
flat $\Lambda$CDM model: $\Omega_{m0}=0.25$; $\sigma_8=0.8$;
$n_{s}=1.0$; $h=0.72$, $\Omega_{\rm b,0}=0.04$. The transfer function
for the simulations was generated using the {\tt cmbfast} code
\citep{SeljakZaldarriaga1996}. The initial conditions were lain down
at redshift $z=49$ using the code {\tt 2LPT}
\citep{Scoccimarro1998,Crocceetal2006}. Each initial condition was
integrated forward using the publicly available cosmological $N$-body
code {\tt Gadget-2} \citep{Springel2005}. Snapshots of the phase space
were captured at 11 logarithmically-spaced intervals between $a=0.5$
and $a=1.0$.

In order to generate full-sky nonlinear ISW maps we roughly follow the
strategy described in \citet{Caietal2010}, but with some minor
changes. Full details of how we construct our maps can be found in
Appendix~\ref{app:ISW}.  In summary, we used the density and
divergence of momentum fields to solve for $\dot\Phi$ for each
snapshot. We then constructed a backward light-cone from $z=0.0$ to
$z=1.0$ for $\dot{\Phi}$. We then pixelated the sphere using the {\tt
  HEALPix} equal-area decomposition, taking the pixel resolution to be
$N_{\rm side}=256$, which corresponds to 786,432 pixels on the
sphere. For each pixel location, we then fired a ray through the past
light-cone of $\dot\Phi$ and accumulated the line-of-sight integral
given by \Eqn{eq:ISW}.  Note that we only consider the ISW signal
coming from $z<1$, since we do not expect a significant
cross-correlation between the relatively low-redshift density slices
for SDSS and the ISW from $z>1$. The top panel of
Fig.~\ref{fig:ISWMap} shows one of the ISW maps that we have
generated from the {\tt zHORIZON} simulations.

We next generated the projected density maps. These were done by first
constructing the projected density map associated with each snapshot
$a_i$. The density field for a given snapshot was obtained as follows.
To each snapshot $a_l$ we associate a specific comoving shell
$[\chi_{l-1/2},\chi_{l+1/2}]$ (see Appendix~\ref{app:ISW} for more
details). We then select all of the particles that fall into the shell
for that epoch, i.e. the $i$th dark matter particle in the $a_l$th
snapshot, is accepted in the shell if $\chi_{l-1/2}<|\bx_{\rm
  i}-\bx_{\rm O}|\le\chi_{l+1/2}$, where $\bx_{\rm i}$ and $\bx_{\rm
  O}$ are the coordinates of the $i$th dark matter particle and the
observer, respectively.  Note that if a given value of $\chi$ is
larger than $L/2, 3L/2, 5L/2, \dots$, then we apply periodic boundary
conditions to produce replications of the cube to larger distances. If
the particle is accepted, then we compute the angular coordinates
$(\theta,\phi)$ for the particle, relative to the observer. Given
these angular coordinates, we then find the associated {\tt HEALPix}
pixel and increment the counts in that pixel. The bottom panel of
Fig.~\ref{fig:ISWMap} shows the projected overdensity map for a thin
redshift shell centred on $z=0.3$. We note that it is hard, by eye, to
note any apparent correspondence between the overdensity and
temperature maps.

At the end we have 11 density maps between $z=1.0$ and $z=0.0$ that
form concentric shells around the observer. These shells were then
co-added using the weights given by our analytic fit to the redshift
distribution of superclusters and supervoids (recall the solid line in
the left panel of Fig.~\ref{fig:s2n}). The resulting co-added
all-sky density maps are then smoothed from $N_{\rm side}=256$ to
N$_{\rm side}=32$, and the positions of the $2n$, $n$ and $n/2$ most
under- and over-dense pixels on this map are recorded. The number $n$
corresponds to a number density of extrema that is identical to that
of real voids and superclusters under the footprint of SDSS DR6.  Each
of those extreme pixels on the $N_{\rm side}=32$ map is then projected
back to the $N_{\rm side}=256$ map on a subset of 64 higher-resolution
pixels, out of which the position of the most under- or over-dense
pixel is used as the target of the AP filter.

%%%%%%%%%%%%%%%%%%%%%%%%%%%%%%%%%%%%%%%%%%%%%%%%%%%%%%%

\begin{figure*}
\includegraphics[width=17cm]{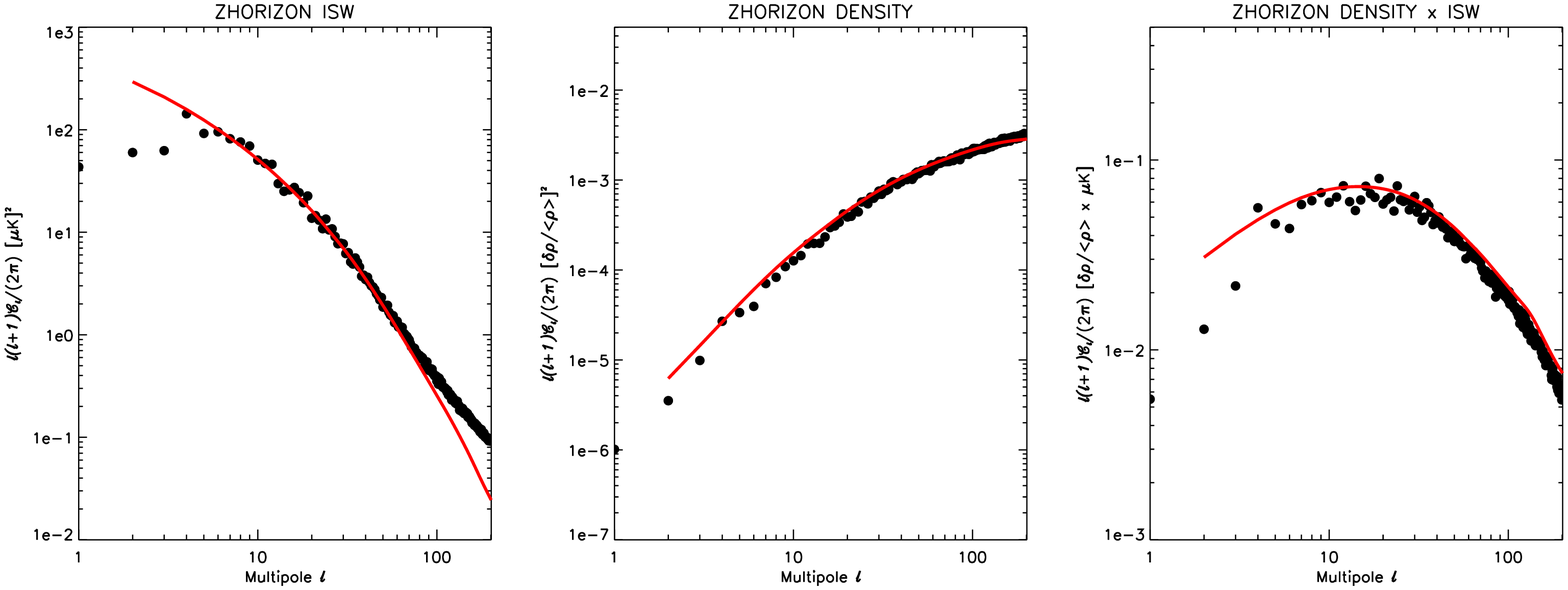}
  \caption{{\it (Left)} Comparison of the average angular power spectra
    obtained from the {\tt zHORIZON}-derived ISW maps (black circles)
    and the theoretical expectation (red solid line). The lack of
    low-$k$ modes causes a low bias in the low multipole range, and
    non-linear evolution introduces some visible power excess on the
    small scales. {\it (Middle)} Comparison of the average angular
    power spectra from the {\tt zHORIZON}-derived density contrast
    maps (black circles) with the linear prediction (red solid line),
    for the redshift range $z\in [0.1,1]$.  The agreement with theory
    is again good on intermediate and small scales, but there seems to
    be again some slight low bias on the largest scales/lowest
    multipoles. {\it (Right)}  Comparison of the average cross angular power spectra from the {\tt zHORIZON}-derived density and ISW maps and the linear theory prediction (red solid line), for the redshift range $z\in[0.1,1.]$. Data from the simulations compare low to theoretical expectations on the large scales/low multipoles due to the finite box size, on the high multipoles due to non-linear evolution, and also show a slight ($\sim 10$ per cent) power deficit on intermediate multipoles due to the finite discretization of the line of sight into 11 shells.   }
\label{fig:cls_zhor}
\end{figure*}

%%%%%%%%%%%%%%%%%%%%%%%%%%%%%%%%%%%%%%%%%%%%%%%%%%%%%%%

\begin{figure}
\includegraphics[width=8cm]{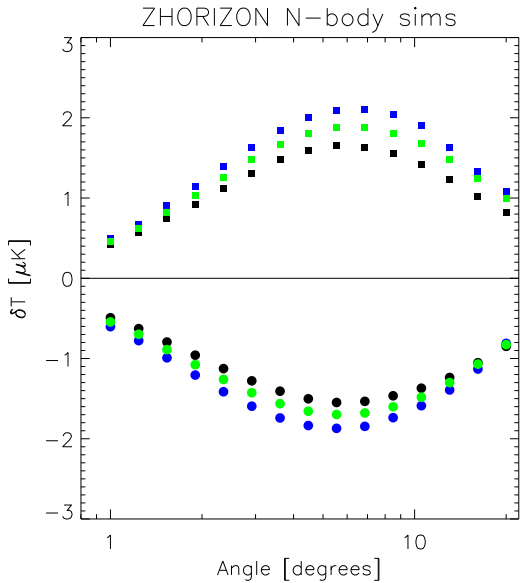}
  \caption{AP filter outputs for ISW maps derived from the {\tt
      zHORIZON} simulations. Black, green and blue symbols refer to
    density excursions in the projected density maps having twice,
    same and half the angular number density as voids and
    superclusters from G08. Circles correspond to under-densities and
    squares to over-densities. These excursions were identified in
    projected density maps smoothed down to pixels of $\sim 2\,\deg$
    on a side (N$_{\rm side}=32$).}
\label{fig:zh_sims}
\end{figure}

%%%%%%%%%%%%%%%%%%%%%%%%%%%%%%%%%%%%%%%%%%%%%%%%%%%%%%%

\subsection{Validation tests of {\tt zHORIZON} derived maps}
\label{sec:zhor_results}

Before we apply the analysis methods of G08 to our ISW and density
maps, we first test the consistency of the maps themselves. To do
this, we compute the angular auto-power spectra of the ISW temperature
maps for each of the 8 {\tt zHORIZON} runs. The left panel of
Fig.~\ref{fig:cls_zhor} presents the ensemble-averaged temperature
power spectrum for the ISW effect. The measurements from the
simulations are represented by the solid black points. The prediction
from linear theory is given by the solid red line. For multipoles in
the range $5<\ell<70$, the agreement between the two is excellent. At
low multipoles ($\ell < 5$) the absence of power in the simulation on
scales larger than $L$ induces a low bias. Instead, on scales
$\ell>70$, the non-linear evolution of potentials substantially boosts
the signal relative to linear by means of the Rees--Sciama effect.

We next compute the angular auto-power spectrum of the projected
density contrast maps, with the projection extending from $z=0.1$ up
to $z=1$ (middle panel of Fig.~\ref{fig:cls_zhor}). In this case,
the projected density angular power spectrum shows good agreement with
the linear-theory expectations in the intermediate- and high-multipole
range ($5<\ell<200$), and some hints for power deficit on the large
scales/low $\ell$-s, which would probably be due to the lack of $k$
modes beyond the box size of the simulations. 

 Finally, the right panel of Fig.~\ref{fig:cls_zhor} compares the ISW
-- density cross-correlation estimated from the simulations with the
linear theory.  As for the other cases, there exists some power
deficit at low multipoles due to finite volume effects in the
simulations.  At high multipoles the linear theory prediction lies
above the simulations, this owes to the fact that in the deeply
non-linear regime potentials do not decay, but grow with time through
the Reese-Sciama mechanism, and hence this leads to a suppression of
power \citep[for further details see][]{Smithetal2009}. On
intermediate angular scales the theoretical prediction is roughly
$\sim 10$ per cent higher than the simulations, although with significant
scatter. This slight mismatch is likely due to the construction of the
weighted projected density field, since the ISW auto-spectra are in
excellent agreement with the simulations.

%%%%%%%%%%%%%%%%%%%%%%%%%%%%%%%%%%%%%%%%%%%%%%%%%%%%%%%

\subsection{Aperture analysis of the {\tt zHORIZON} maps}
\label{sec:zhor_results}

Having validated the simulated maps, we now repeat the AP
analysis. Since here we have both full-sky density and temperature
maps, we prefer not to apply any sky mask. Thus, these predictions
will not be affected by incomplete sky coverage. We repeat the steps
described in \S\ref{sec:Granett} for finding the locations of the
density peaks/troughs in each of the simulated smoothed maps. Then, as
before, we apply the AP filters to the selected centroid positions for
the 8 ISW maps from the {\tt zHORIZON} simulations.

Figure ~\ref{fig:zh_sims} presents the results from this analysis. The
square and circular symbols denote the results for our effective
supercluster and supervoid regions, respectively. The blue, green and
black colours correspond to the set of extreme pixels which have half,
equivalent and double the angular number density of the real
supervoids and superclusters found in G08's analysis. We see that the
peak of the average AP output has a temperature of the order
2\,$\mu$K, and occurs for apertures of scale $\sim6$--$7\degr$.

On comparison with the predictions from our Gaussian realisations, we
find that the fully nonlinear ISW maps are in close agreement
(c.f.\S\ref{ssec:gaussian}). The are however small differences.  The
peak signal is shifted to slightly larger scales for the full
nonlinear case. Also, the shape of the curves obtained from the {\tt
  zHORIZON} simulations appears smoother than in the Gaussian
simulation case, which shows dips and troughs that are absent in
 Fig.~\ref{fig:zh_sims}. We believe that these small differences are
likely a consequence of the fact that the Gaussian realisations include
intrinsic CMB noise and possess a sky-mask. 

Actually,  after applying the real sky masks to the simulated maps, we find
  that the peak amplitudes and the general shapes of the functions in
  Fig.(\ref{fig:zh_sims}) become distorted at the $\sim$10 per cent level.

However, the most important point to note, is that on angular scales of the order
3--4\degr, the Gaussian and fully nonlinear simulations are in close
agreement: the difference induced by adopting a slightly different
cosmological model should introduce changes in the ISW amplitude at
the 2 per cent level, and the ISW generated beyond $z=1$ seems to have
little impact as well.  We thus conclude this section by noting that
that the excess temperature signal found by G08, and now confirmed by
us in \S\ref{sec:Granett}, appears to be incompatible with the
evolution of gravitational potentials in the standard $\Lambda$CDM
model. Our results from both Gaussian realisations and ISW maps derived from the N-body simulations are in agreement with \citet{Granettetal2008c}, who found no signature at the few degree scale on voids and clusters with the amplitude found on real WMAP data when producing an ISW map out of the distribution of Luminous Red Galaxies in Sloan data. 

%%%%%%%%%%%%%%%%%%%%%%%%%%%%%%%%%%%%%%%%%%%%%%%%%%%%%%%

\section{Significance, Systematics and alternative models}\label{sec:discussion}

%%%%%%%%%%%%%%%%%%%%%%%%%%%%%%%%%%%%%%%%%%%%%%%%%%%%%%%

\subsection{Estimating the significance}

The direct comparison of  Fig.~\ref{fig:wmapdata} and
\ref{fig:gsims} reveals clear differences between the observed data
and theoretical predictions. Not only is the amplitude of the maximum
signal in the real data a factor of $\sim$5 times larger than the
average in the Gaussian realisations, but the dependence of the signal
on the filter scale shows a different shape. More quantitatively, the
results from the W-band WMAP data for an aperture size of 3.6\degr, are of
the order $\sim3.4\,\sigma$ away from the supervoid simulation
average, and $\sim2.1\,\sigma$ away from the average for the case of
superclusters. In terms of probability, for an aperture scale of
3.6\degr, only 5 out of the 5000 realisations possessed an ISW
signal, from supervoid regions, with a temperature decrement lower
than the one found in real data, and 97 of the realisations for
superclusters exceeded the value obtained for the real data.  Taken at
face value, this analysis seems to exclude the Gaussian $\Lambda$CDM
hypothesis at $\sim 4\sigma$ significance. However, this is an {\em a
  posteriori} estimate, since we have neglected the fact that we also
looked for a signal at other aperture scales.

If, for the W-band WMAP data, we include the measurements from the 15 different angular aperture
scales between 1--20\degr and take into account their
covariance, then the significance drops. Under the assumption of
Gaussian statistics, we find that the WMAP outputs for voids produce a
$\chi^2_{\rm voids}=23.5$ ($n_{\rm dof}=15$). The corresponding figure
for superclusters is $\chi^2_{\rm superclusters}=26.6$ ($n_{\rm
  dof}=15$). If we treat these two constraints as being independent
then their combination yields $\chi^2_{\rm both}=50.0$ ($n_{\rm
  dof}=30$). In terms of probability, this means that the WMAP data
have a 0.012 probability (i.e. $<2$ per cent chance or $\sim$2.2$\,\sigma$ under Gaussian statistics) 
of being consistent with the evolution of gravitational potentials in the $\Lambda$CDM
model. If we consider the null hypothesis of no ISW signatures expected at all (for which stacking on voids and superclusters should leave no temperature decrement/increment), then the results lie at $2.6\,\sigma$ away from this scenario.

We next study the dependence of the statistical significance on the number of substructures considered in the analysis. While the original catalogue of voids and superclusters of G08 contains 50 entries, we now repeat our tests after considering two subsamples containing only the first 30 and 40 objects. For these subsamples, the adopted values of the Gaussian threshold were $\nu=$ 2.87 and 2.97. In these cases, the pattern found for the full catalogue is reproduced: the stacked voids give a temperature decrement of $\lesssim -10$\,$\mu$K, at $\sim 3$--$3.5$\,$\sigma$;  the stacked superclusters give a temperature increment of $\gtrsim 7$\,$\mu$K, at the level of $\sim 2.2$--$2.4$\,$\sigma$. On comparison with Gaussian realisations, we find that, after considering all aperture radii, results for the first 40 superstructures are in lower tension with the outputs of Gaussian realisations (at the level of $<3.0$ per cent or $1.9$\,$\sigma$). This level of tension further decreases when considering only the first 30 superstructures ($\sim 27$ per cent or $0.6\,\sigma$), showing that the tension of WMAP data wrt to Gaussian realisations relaxes as fewer structures are included in the analysis. This is somehow expected from the Gaussian realisations, for which the statistical significance for the ISW increases with decreasing thresholds. This is in apparent contradiction with Table 1 of G08, where it is shown that the statistical significance of their ISW measurement at a scale of 4\degr decreases when increasing the number of structures from 50 ($4.4\,\sigma$) to 70 ($2.8\,\sigma$). In G08 it is argued that by considering more structures one may be diluting the signal by including unphysical structures, an extent that cannot be tested in our Gaussian maps since it is strictly associated to the algorithms identifying voids and superclusters in the galaxy catalogues. 

In summary, according to our Gaussian simulations, the $\sim 4\,\sigma$ deviation wrt $\Lambda$CDM expectations found at $\sim 4$\degr  aperture radius decreases to $\sim 2.2\,\sigma$ when including different filter apertures in the range $[1\degr,20\degr]$, and lies, in this case, $2.6\,\sigma$ away from the null (no ISW) case. While this tension relaxes when considering fewer structures, the significance of the detected signal seems to decrease when considering more than 50 voids and superclusters (see Table ~1 of G08), in an opposite trend to what is suggested by our Gaussian ISW realisations.

We conclude that most of the significance of the G08 result is at odds with ISW $\Lambda$CDM  predictions, both in amplitude and scale/aperture radius dependence, and that this tension considerably reduces when more aperture radii and structure sub-samples are considered in the analysis. 

%%%%%%%%%%%%%%%%%%%%%%%%%%%%%%%%%%%%%%%%%%%%%%%%%%%%%%%

\subsection{Tests for systematics}

%%%%%%%%%%%%%%%%%%%%%%%%%%%%%%%%%%%%%%%%%%%%%%%%%%%%%%%

\begin{figure}
\includegraphics[width=8cm]{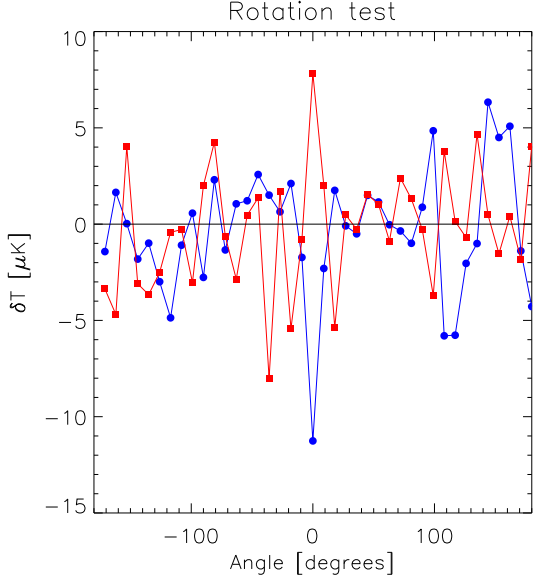}
\caption{Rotation test on AP filter outputs for an aperture size of
  3.6\degr . Blue circles (red squares) correspond to voids
  (superclusters).}
\label{fig:rottest}
\end{figure}

%%%%%%%%%%%%%%%%%%%%%%%%%%%%%%%%%%%%%%%%%%%%%%%%%%%%%%%

Given the high level of discrepancy existing between the pattern found at $3.6$\degr aperture radius and  ISW $\Lambda$CDM expectations, we next test the possibility of systematics in WMAP data giving rise to the observed signal.

\begin{itemize}

\item {\bf Rotation test:} We first conduct a rotation test assessing
  the statistical significance of the AP outputs for a 3.6\degr
  scale. In this way, we probe the possibility of any other signal
  (apart from CMB) contributing to the uncertainty of the AP filter
  outputs and hence modifying the statistical significance found for
  WMAP data.  We rotate in galactic longitude (in steps of 9\degr) the
  AP filter targets with respect to the real positions of supervoids
  and superclusters. In the absence of systematics, this should
  provide AP outputs compatible with zero.

Figure~\ref{fig:rottest} presents the results from this analysis.  The
blue circular symbols represent the supervoid regions and the red
square symbols denote the superclusters. At zero rotation lag we
clearly obtain a signal of higher amplitude than in any other rotation
bin. We have verified that this signal does not arise as a consequence
of a small subset of the supervoids/superclusters. Instead, the signal
is approximately evenly distributed among all structures. From the
sample of rotated bins only, the estimated significance for the
3.6\degr aperture is 4.1\,$\sigma$ for voids, 2.7\,$\sigma$ for
superclusters and 3.8\,$\sigma$ combined. This is within 1--$\sigma$
from the significance levels obtained with the Gaussian realisations.

%%%%%%%%%%%%%%%%%%%%%%%%%%%%%%%%%%%%%%%%%%%%%%%%%%%%%%%

\begin{figure}
\includegraphics[width=8cm]{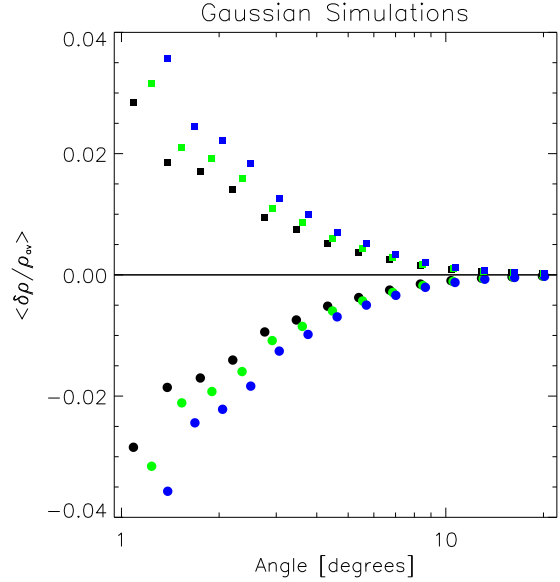}
  \caption{Pattern induced versus AP filter scale by a contaminant
    following the local matter density.}
\label{fig:pointsource}
\end{figure}

%%%%%%%%%%%%%%%%%%%%%%%%%%%%%%%%%%%%%%%%%%%%%%%%%%%%%%%

\begin{figure*}
\centering{
\includegraphics[width=8cm,clip=,angle=90]{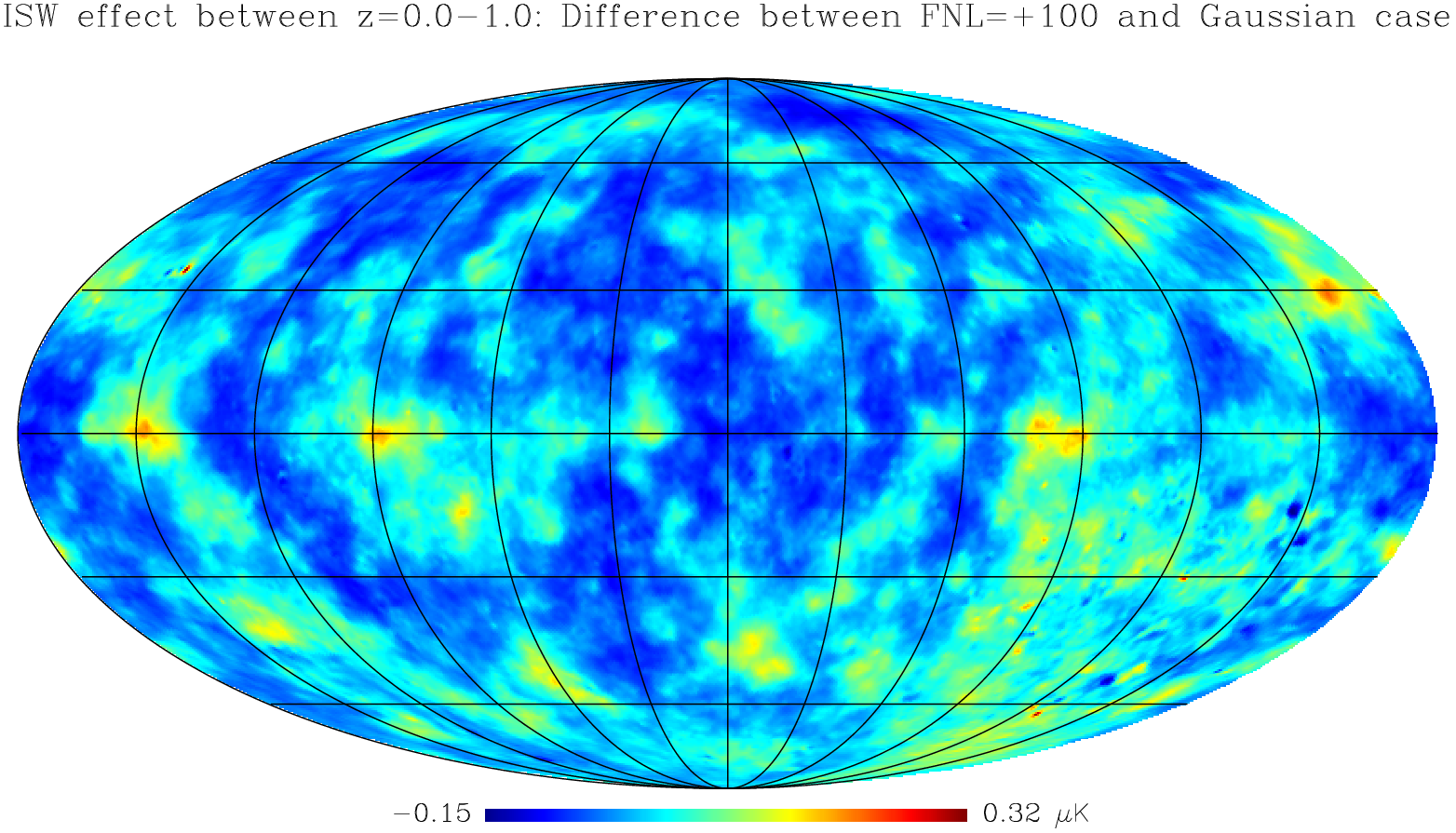}
\includegraphics[width=8cm,clip=,angle=90]{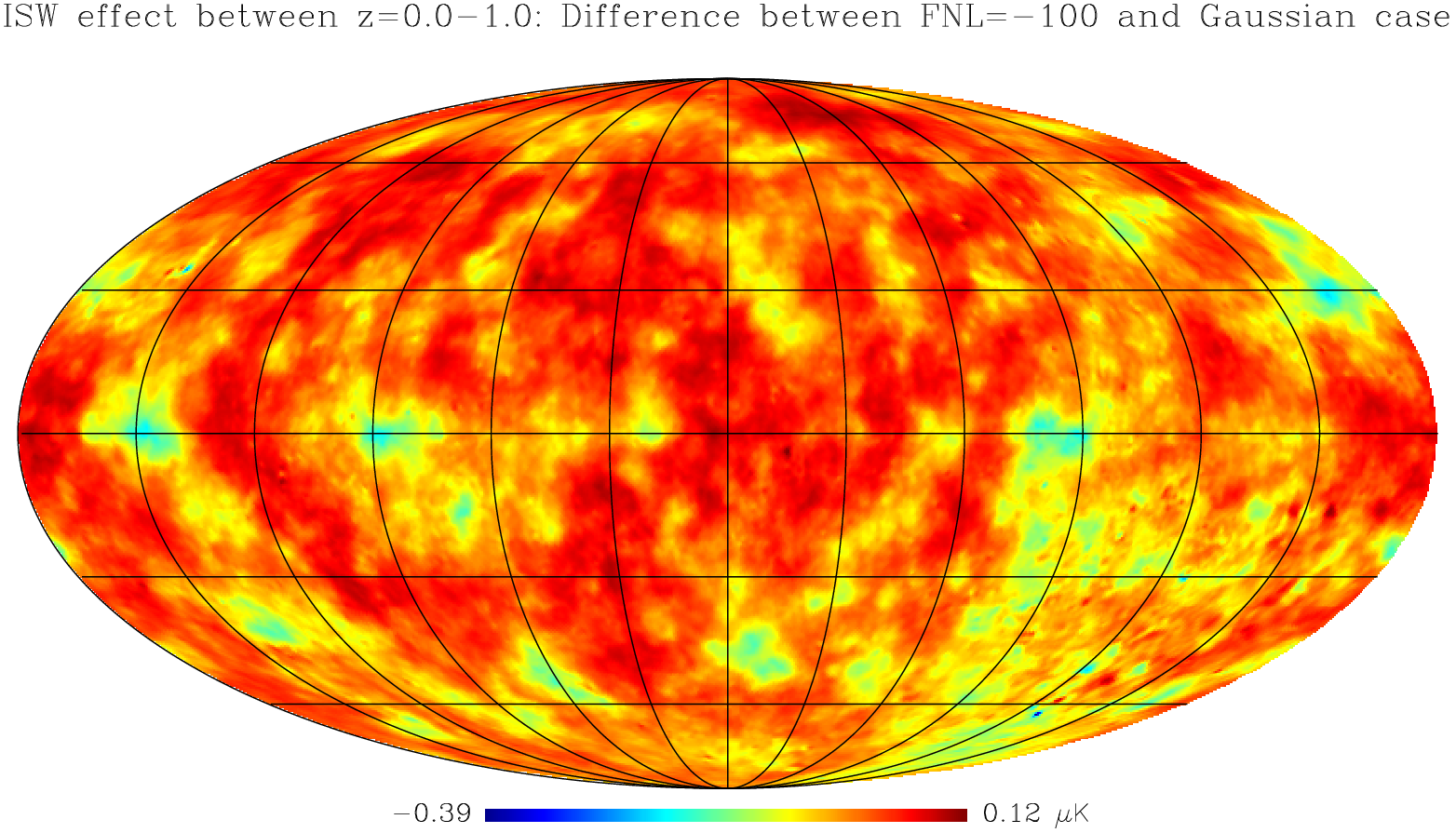}}
\caption{\small{Relative difference between full-sky ISW temperature
    maps for models with primordial non-Gaussianity and Gaussian
    models for flat $\Lambda$CDM universes. {\em Top panel:} presents
    $\Delta T^{\rm ISW}(\FNL=+100)-\Delta T^{\rm ISW}(\FNL=0)$. {\em
      Bottom panel:} presents the same but for $\Delta T^{\rm
      ISW}(\FNL=-100)-\Delta T^{\rm ISW}(\FNL=0)$. All maps were
    smoothed with a Gaussian filter of $\rm FWHM=1.0\degr$, which
    roughly corresponds to the scales at which the signal for the AP
    filters peaks. The large-scale temperature fluctuations appear
    slightly cooler/hotter in Universes with $\FNL$
    positive/negative.}
\label{fig:fnl}}
\end{figure*}

%%%%%%%%%%%%%%%%%%%%%%%%%%%%%%%%%%%%%%%%%%%%%%%%%%%%%%%

\item {\bf Density-dependent contaminant:} Another possible systematic
  is some combination of contaminants present in the superclusters
  that {\em increases} the emission in these structures in a frequency
  independent manner. This could happen if the CMB cleaning algorithms
  remove only the frequency varying part of the contaminant signal,
  but leave behind a DC level. In such a scenario, one might assume
  that the signal is linearly responding to fluctuations in the
  projected matter density.  We incorporate this effect into our
  Gaussian realisations by substituting the total CMB map by a signal
  that is proportional to the projected matter density field in our
  Gaussian realisations. This should provide us with some idea of the
  scale-dependence for a contaminant of this nature.

Figure~\ref{fig:pointsource}, presents the results from this exercise.
The symbol styles and point colours are the same as in
Fig.~\ref{fig:gsims}. The recovered shape resembles that for a
standard matter auto-correlation function, and is remarkably different
from that of Fig.~\ref{fig:wmapdata}. Actually, the profile from
real data in Fig.~\ref{fig:wmapdata} seems to be an intermediate
case between the scenario depicted in Fig.~\ref{fig:pointsource} and
the theoretical predictions of Fig.~\ref{fig:gsims}. However, if
most of the observed amplitude at 3.6\degr ($\sim 8-2 = 6\,\mu$K out
of the total $\sim 8\,\mu$K observed) is to be caused by this type of
contaminant, then the scale dependence of the output should
accordingly be much closer to the one shown in
Fig.~\ref{fig:pointsource}, and this is not the case. Note that we
have assigned $\sim 2\,\mu$K to ISW in this estimation. If
contaminants in the position of the AP targets were Poisson
distributed, then the profiles obtained in
Fig.~\ref{fig:pointsource} would approach zero faster as aperture
radii increase, yet in stronger disagreement with
Fig.~\ref{fig:wmapdata}.

\item {\bf Superstructure selection effects:} It might be argued that
  our identification of under- and over-dense regions in our simulated
  density maps does not match well the selection of supervoids and
  superclusters in the G08 data. Whilst in principle both approaches
  should generate the most under- and over-dense regions in volume
  limited samples of the universe, differences in the exact details of
  the {\tt ZOBOV} and {\tt VOBOZ} implementations may introduce subtle
  but important differences in the selection of regions. This would
  have an impact in the final definition and selection of targets for
  the AP filters. Properly addressing this issue would require a
  thorough implementation of the {\tt ZOBOV} and {\tt VOBOZ}
  algorithms to semi-analytic galaxy samples embedded in our
  simulations (and this goes beyond the scope of the current
  work). Having said this, the stability of our results with respect
  to the actual density peak threshold adopted, suggests that this
  would not critically affect our conclusions.

\end{itemize}

%%%%%%%%%%%%%%%%%%%%%%%%%%%%%%%%%%%%%%%%%%%%%%%%%%%%%%%

\subsection{Primordial non-Gaussianity}

In this section we explore whether the G08 result is compatible with a
non-Gaussian distribution for the primordial potential perturbations.
In particular, we consider the well known local model for primordial
non-Gaussianity, characterized by a quadratic correction to the Gauge
invariant Bardeen's potential perturbation \citep[see][and references
  therein]{Komatsuetal2011short}:
\be 
\phiNG(\bx)=\phiG(\bx)+
\FNL \left[\phiG(\bx)^2-\left<\phiG^2(\bx)\right>\right]\ ,
\label{eq:fnl}
\ee
where $\phiG(x)$ is the Gaussian potential perturbation after matter
radiation equality, scaled in terms of units of $c^2$ to yield a
dimensionless quantity. Following standard convention,
$\phiNG(\bx)\equiv-\Phi^{\rm Newton}(\bx)$ (i.e. the $\Phi$ in
\Eqn{eq:ISW}). The term $\left<\phiG^2(x)\right>$ is subtracted to
ensure that $\phiNG$ is a mean zero field. In linear theory the
typical fluctuations are of the order $\phiNG\sim10^{-5}$, and so the
non-Gaussian corrections are of the order \mbox{$\sim0.1\%
  (\FNL/100)(\phiG/10^{-5})^2$}.

In order to explore the observable consequences of such a
modification, we have generated a set of simulated ISW maps with
Gaussian initial conditions, i.e. $\FNL=0$, and with non-Gaussian
initial conditions $\FNL=\{+100,-100\}$.  These maps were generated
from $N$-body simulations seeded with Gaussian and non-Gaussian
initial conditions following the methodology of Appendix\ref{app:ISW}.
The simulations that we employ were fully described in
\citet{Desjacquesetal2009}. In brief, these were performed using {\tt
  Gadget-2}, and followed $N=1024^3$ dark matter particles in a box of
size $L=1600\Mpc$. The cosmological model of the simulations was
consistent with the WMAP5 data \citep{Komatsuetal2009short}. We use a
sub-set of these simulations that were used elsewhere for
gravitational lensing analysis \citep{Marianetal2011,Hilbertetal2012}.
The simulations were set up to have the same initial random phases for
all three models, this enables us to cancel some of the cosmic
variance and so permit us to better explore the model differences.

The top panel of Fig.~\ref{fig:fnl} presents the differences between
the ISW temperature maps in a Universe with $\FNL=+100$ and
$\FNL=0$. The bottom panel shows the same but for the case $\FNL=-100$
and $\FNL=0$. Note that all of the maps were smoothed with a Gaussian
filter of ${\rm FWHM}=1\degr$ before being differenced. Note also that
we have only included the ISW contributions between $z=0.0$ and
$z=1.0$. Clearly, the presence of primordial non-Gaussianities can
induce shifts in the temperatures of the peaks and troughs of the
distribution. However, these shifts are modest for $\FNL=\pm100$,
leading to changes that are $<1\mu {\rm K}$. This suggests that values
of $\FNL$ on the order of $\sim1000$, might be able to explain the AP
analysis of the WMAP results. However, such large values for $\FNL$
would be grossly inconsistent with the values of $\FNL$ obtained from
the CMB temperature bispectrum, which currently gives
\mbox{$-10<f_{\rm NL}^{\rm local} <74$} \citep[95\%
  C.L.,][]{Komatsuetal2011short}. It therefore seems unlikely that the
scale-independent local model of primordial non-Gaussianity is the
correct explanation for the excess signal.

%%%%%%%%%%%%%%%%%%%%%%%%%%%%%%%%%%%%%%%%%%%%%%%%%%%%%%%

\section{Conclusions}
\label{sec:conclusions}

In this work we have studied the imprint of superclusters and
supervoids in the temperature map of the CMB from the WMAP experiment.
Our work further explores the signature first detected in
\citet[][G08]{Granettetal2008b}.

In \S\ref{sec:theory} we theoretically showed that if G08 had applied
a standard angular cross-power spectrum analysis of the
superstructures they found in the SDSS LRG data, then the expected
significance for a $\Lambda$CDM model should have been $<1.5\,\sigma$.

In \S\ref{sec:Granett} we cross-checked the G08 analysis directly and
found identical conclusions: on scales $\sim3.6\degr$ there was a
$\sim4\sigma$ detection significance for excess signal associated with
the supervoids and superclusters.

In \S\ref{sec:tests} we performed a series of tests exploring whether
these findings are consistent with the standard $\Lambda$CDM model.
Gaussian Monte-Carlo realisations of the ISW effect and the LSS were
unable to produce such large signals. We then investigated whether
this was a consequence of our simplified Gaussian realisations. We did
this by generating fully non-linear ISW maps from large volume
$N$-body simulations. These simulated maps confirmed the findings of
the simpler Gaussian realisations.

In \S\ref{sec:discussion} we used the Gaussian Monte-Carlo realisations
to explore the significance of the deviations from the $\Lambda$CDM
model found in the WMAP data. We found that for aperture photometry
analysis of the maps on scales $3.6\degr$, results from WMAP data
are lying about $\sim 4\,\sigma$ away 
from $\Lambda$CDM expectations. However, on taking into account the 15
aperture scales examined, the significance of the discrepancy dropped
to $<2$ per cent chance ($2.2\,\sigma$) of the result being consistent with the $\Lambda$CDM
model. In this case, results remained $2.6\,\sigma$ away from the null (ISW-free) scenario where structures leave no signatures on the CMB at the linear level. When including fewer structures in the analysis, the tension dropped further, and results for only 30 voids and superclusters were compatible both with  $\Lambda$CDM expectations (at $0.6\,\sigma$) and the null (no ISW) scenario (at $1.3\,\sigma$). Our simulations also suggested that the ISW significance should increase when more structures were included in the analysis, in apparent contradiction with the findings of G08. Hence, most of the detected signal appeared associated to the full set of 50 superstructures and an aperture radius of $3.6\degr$.

We investigated whether the observed pattern at a radius of $3.6\degr$ could be caused by a systematic error in the cleaning of foregrounds in the WMAP data. We found that if the
signal were to be caused by an approximately frequency-independent
emission tracing the density field, then the resulting angular
dependence would be very different to the measured shape found in the
WMAP data.

We next explored whether the observed signal at  $3.6\degr$
 could be generated by primordial non-Gaussianities. We considered the local model, characterized by a
quadratic correction to the primordial potential perturbations, with
the coupling parameter $\FNL$. We found that, for $\FNL$
positive/negative, asymmetric shifts in ISW temperature maps arise.
However, for the values of $\FNL=\pm100$, the changes were $<1\mu$K
(after smoothing the maps down to degree scales). Thus values of
$\FNL$ an order of magnitude higher would be required to explain the
G08 result, and they would be clearly inconsistent with current
constraints on $\FNL$ from WMAP.

It is possible that the G08 result may also be explained by other more
exotic scenarios, e.g., non-Gaussianity arising from the presence of a
non-zero primordial equilateral or orthogonal model bispectrum (a
consequence of non-standard inflationary mechanisms); alternatively it
might arise as a direct consequence of modifications to Einstein's
general theory of relativity, \citep{JainKhoury2010}.  However, more
conservative scenarios involving some combination of artifacts and/or
systematics cannot yet be fully discarded.

In the future, we will look with interest to the results from the
{\it Planck} satellite as to whether this signal represents a data artifact,
or whether it constitutes a genuine challenge to the $\Lambda$CDM model
and a window to new cosmological physics.

%%%%%%%%%%%%%%%%%%%%%%%%%%%%%%%%%%%%%%%%%%%%%%%%%%%%%%%

\section*{Acknowledgments}\label{lastpage}

It is a pleasure to acknowledge Ra\'ul Angulo, Jose Mar\'\i a Diego,
Marian Douspis, Benjamin Granett, S.~Illic and Istv\'an Szapudi for
useful discussions. We also thank Laura Marian for carefully reading
the manuscript. We kindly thank Vincent Desjacques for providing us
with access to his non-Gaussian realisations. C.H-M. is a Ram\'on y Cajal fellow of the
Spanish Ministry of Economy and Competitiveness. The work of RES was supported by Advanced Grant 246797 `GALFORMOD' from the European Research Council. We acknowledge the use of
the {\tt HEALPix} package 
\citep{Gorskietal2005} and the {\tt LAMBDA} data base. We thank
Volker Springel for making public his code {\tt Gadget-2}, and Roman
Scoccimarro for making public his {\tt 2LPT} code. We acknowledge the
ITP, University of Z\"urich for providing assistance with computing
resources. 
    
\bibliographystyle{mn2e}
\bibliography{refs}

%%%%%%%%%%%%%%%%%%%%%%%%%%%%%%%%%%%%%%%%%%%%%%%%%%%%%%%

\appendix

%%%%%%%%%%%%%%%%%%%%%%%%%%%%%%%%%%%%%%%%%%%%%%%%%%%%%%%

\section{Full sky ISW maps from $N$-body simulations}\label{app:ISW}

In this section we aim to construct full-sky ISW maps using a suite of
$N$-body simulations. Our approach is similar to that described in
\citet{Caietal2010}, but with some modifications. To be more precise,
we aim to compute the line-of-sight integral \Eqn{eq:ISW}, but taking
into account the full nonlinear evolution of $\dot{\Phi}$. The steps
we take to achieve this are described below.

%%%%%%%%%%%%%%%%%%%%%%%%%%%%%%%%%%%%%%%%%%%%%%%%%%%%%%%

\subsection{Determining $\dot{\Phi}$}\label{app:detphidot}

In order to obtain $\dot{\Phi}$ directly from the $N$-body
simulations, we make use of \Eqn{eq:ISW1}, which tells us that our
desired quantity can be determined from knowledge about $\delta(\bk,a)$
and $\dot{\delta}(\bk,a)$.  In simulations, measuring $\delta(\bk,a)$
is relatively straightforward, whereas its time derivative is more
complicated. As was shown by \citet{Seljak1996a}, this latter quantity
may be obtained from the perturbed continuity equation
\citep{Peebles1980}:
\be {\bf\del}\cdot\left[1+\delta(\bx;t)\right]{\bf v}_p(\bx;t)=-
a(t)\dot{\delta}(\bx;t) \ , \ee
where ${\bf v}_p(\bx;t)$ is the proper peculiar velocity field. If we
define the pseudo-peculiar momentum field to be,
\be {\bf p}(\bx;t)\equiv\left[1+\delta(\bx;t)\right]{\bf v}_p(\bx;t)\ , \ee
then in Fourier space we may solve the continuity equation directly to
find
\be \dot{\delta}(\bk;t)= i\bk\cdot{\bf p}(\bk;t) /a(t)\ .\ee
Hence, our final expression becomes,
\be \dot{\Phi}(\bk;t) = \Fka \left[
  \frac{H(t)}{a(t)}\delta(\bk;t)-\frac{i\bk\cdot{\bf
      p}(\bk;t)}{a^2(t)} \right] \ , \label{eq:pdot}\ee
where we defined the function 
\be {\mathcal F}(k) \equiv \frac{3}{2}\Omega_{\rm m0}
\left(\frac{H_0}{k}\right)^2  \ . \ee
Thus in order to estimate $\pdot$, we simply require estimates of both
the density field and pseudo-peculiar momentum field in Fourier
space. 

The dark matter density field in an $N$-body simulation can be written
as a sum of Dirac delta functions,
\be \rho(\bx)= \sum_{l=1}^{N} m_l \delta^D(\bx-\bx_l) \ ,\ee
where $N$ is the number of particles and $m_l$ is the mass of the
$l$th particle, and we take all particles to have equal mass. The
density field averaged on a cubical lattice can then be obtained
through the convolution,
\ba
\rho_{g}(\bx_{ijk}) 
& = & 
\frac{1}{V_{W}}\int \dx \rho(\bx) W(\bx_{ijk}-\bx) \ ; \nonumber \\
& = &
\frac{m}{\Vu}\sum_{l}^{N} W(\bx_{ijk}-\bx_l) \ ,
\ea
where $\bx_{ijk}$ labels the lattice point, $W$ represents the
dimensionless window function of the mass assignment scheme. This
window function is normalized such that \mbox{$V_{W} = \int \dx'
W(\bx-\bx')$}. The filter function $W$ that we adopt throughout is the
`cloud-in-cell' charge assignment scheme
\citep{HockneyEastwood1988}. Hence, our estimate for the density
fluctuation is
\ba 
1+\widehat{\delta(\bx_{ijk})}
 & = & 
\frac{1}{N}\frac{\Vu}{V_{W}} \sum_{l}^{N} W(\bx_{ijk}-\bx_l)  \ ,\nonumber \\
& = & 
\frac{N_{\rm cell}}{N} \sum_{l}^{N} W(\bx_{ijk}-\bx_l)  \ ,
\ea
where $N_{\rm cell}=\Vu/V_{W}$ is the total number of grid cells.

The pseudo-momentum field may be estimated in a similar fashion.  For
convenience we write,
\be {\bf p} = \left[1+\delta(\bx)\right]{\bf u}(\bx) a(t) \ , \ee
where ${\bf u}={\bf v}_p/a$ is the comoving peculiar velocity
field. The particle momentum field is then written as
\be \left[(1+\delta){\bf u}\right](\bx) = \frac{\Vu}{N}
\sum_{l}^{N}\delta^{D}(\bx-\bx_l) {\bf u}_l \ . \ee
This may be convolved with the mass assignment scheme to obtain the
mesh averaged quantity
\be  \left[(1+\delta){\bf u}\right](\bx_{ijk}) =
\frac{1}{N}\frac{\Vu}{V_{W}} \sum_l^{N} {\bf u}_l W(\bx_{ijk}-\bx_l) \ .\ \ee
Thus our estimate for the pseudo-momentum field is
\be \widehat{\bf p}(\bx_{ijk}) = a(t) \frac{N_{\rm cell}}{N}\sum_l^{N}
    {\bf u}_l W(\bx_{ijk}-\bx_l) \ . \ee

The density Fourier modes were then estimated using the publicly
available {\tt FFTW} routines \citep{FFTW}, and each resulting mode
was corrected for the convolution with the mass-assignment window
function. For the CIC algorithm this corresponds to the following
operation:
\be \delta_{\rm d}(\bk)=\delta_{\rm g}(\bk)/W_{\rm CIC}(\bk) \ ,\ee
where
\be W_{\rm CIC}(\bk)=\prod_{i=1,3}\left\{\left[\frac{\sin 
\left[\pi k_i/2k_{\rm Ny}\right]}{\left[\pi k_i/2k_{\rm Ny}\right]}\right]^2\right\} \ee
and where sub-script d and g denote discrete and grid quantities, and
where $k_{\rm Ny}=\pi N_{\rm g}/L$ is the Nyquist frequency, and
$N_{\rm g}$ is the number of grid cells \citep{HockneyEastwood1988}.

To obtain the real space $\pdot(\bx,t)$, we solved for $\pdot(\bk,t)$
in Fourier space using Eq.~(\ref{eq:pdot}), set the unobservable $k=0$
mode to zero, and inverse transformed back to real space.

%%%%%%%%%%%%%%%%%%%%%%%%%%%%%%%%%%%%%%%%%%%%%%%%%%%%%%%

\subsection{Reconstructing the light-cone for $\dot{\Phi}$}

We now wish to construct the past light-cone for the evolution of
$\dot{\Phi}$, however we only have a finite number of snapshots of the
particle phase space from which to reconstruct this. It is usually a
good idea to space snapshots logarithmically in expansion factor, and
for simplicity we shall now assume that to be true. The light-cone can
then be constructed as follows:

\begin{itemize}

\item 
Find $\dot\Phi(\bx_{ijk},a_l)$ for every output $l$ on a periodic
cubical lattice, using the techniques described in
Appendix~\ref{app:detphidot}.

\item Place an observer at the exact centre of the simulation cube,
  $\bx_{\rm O}$, and compute the comoving distances from the observer
  to the expansion factors, $a_{l-1/2}$, $a_l$, and $a_{l+1/2}$, and
  label these distances $\chi_{l-1/2}$, $\chi_{l}$ and
  $\chi_{l+1/2}$. Here,
\[\log a_{l\pm1/2}=\log a_l \pm\Delta \log a/2\ ,\] 
with $\Delta \log a$ being the logarithmic spacing between two
different expansion factors. The comoving distance from the observer
at $a_0$ to expansion factor $a$ is given by:
\[\chi(a)=\int_{a}^{a_0} \frac{c da}{a^2H(a)} \ .\]
Hence, the intervals $[\chi_{l-1/2},\chi_{l+1/2}]$ form a series of
concentric shells centred on the observer.
\item Construct a new lattice for the $\dot\Phi$ values on the light
  cone. This is done by associating to each snapshot $a_l$, a specific
  comoving shell $[\chi_{l-1/2},\chi_{l+1/2}]$, and taking only those
  values for $\dot\Phi(\bx_{ijk},a_l)$ that lie within the shell:
  i.e. if
  \[\chi_{l-1/2}<|\bx_{ijk}-\bx_{\rm
    O}|\le\chi_{l+1/2}\ ,\]
then $\dot\Phi(\bx_{ijk},a_l)$ is accepted onto the new grid. Note
that if a given value of $\chi$ is larger than $L/2, 3L/2, 5L/2, \dots$, then
we use the periodic boundary conditions to produce replications of the
cube.
\end{itemize}

%%%%%%%%%%%%%%%%%%%%%%%%%%%%%%%%%%%%%%%%%%%%%%%%%%%%%%%

\subsection{Computing the ISW line-of-sight integral}

Having constructed the backward lightcone for $\dot{\Phi}$, we may now
compute the line-of-sight integral for the ISW effect through
\Eqn{eq:ISW}. In fact we use a slightly different form of this
equation by transforming the integration variable from $t$ to $\log a$:
i.e.
\be \frac{\Delta T(\nhat)}{T_0} = {2 \over c^2}\int_{\ln a_{\rm
    ls}}^{\ln a_{0}} d\log a \frac{\dot{\Phi}(\nhat,\chi)}{H(a)}
\ \label{eq:ISWloga} .\ee
The above expression can then be discretized to give,
\be \frac{\widehat{\Delta T}(\nhat)}{T_0} \approx {2 \over c^2}
\sum_{i=0}^{M} \Delta\log a_i \frac{\widehat{\dot{\Phi}}(\nhat,a_i)}{H(a_i)}
\ \label{eq:ISWloga2} \ .\ee
If we take $\Delta \log a$ to be constant, then this becomes,
\be \frac{\widehat{\Delta T}(\nhat)}{T_0} \approx \frac{2\Delta\!\log a}{c^2}
\sum_{i=0}^{M} \frac{\widehat{\dot{\Phi}}(\nhat,a_i)}{H(a_i)}
\ \label{eq:ISWloga2} \ .\ee
In evaluating the above expression, we take the number of intervals to
be as large as desired, but always evaluating $H(a_i)$ exactly through
the expression \citep{Dodelson2003,Weinberg2008},
\be
H^2(a)=H^2_0\left[\Omega_{\Lambda0}+
\Omega_{\rm m0}a^{-3}-(\Omega_{\rm m0}+\Omega_{\Lambda0}-1)a^{-2}\right]\ ,
\ee
which is valid in the matter dominated epoch for the $\Lambda$CDM
model.  Whereas, for $\dot{\Phi}(a_i)$, we employ the light cone
derived form from the previous subsection, but interpolate the 
value off the 3D mesh using a CIC like scheme. That is:
\ba 
\dot{\Phi}(\bx,a)& = & (1-h_{x})(1-h_{y})(1-h_z)\dot{\Phi}(\bx_{i,j,k})\nn\\
& & + (1-h_{x})(1-h_{y})h_z\dot{\Phi}(\bx_{i,j,k+1})+\dots  \nn \\
& & + h_xh_yh_z \dot{\Phi}(\bx_{i+1,j+1,k+1}) )\ ,\ea
where $\{h_x,h_y,h_z\}$ are the $x$-, $y$-, and $z$-coordinate
separations of the evaluation point $\bx$ and the position vector for
lattice point $(i,j,k)$, in units of the inter-lattice separation.

In making this separation of the evolution of $H(a)$ and $\dot{\Phi}$,
we are effectively assuming that $\dot{\Phi}$ evolves very slowly over
the time interval between snapshots. This is a reasonable assumption
on large scales, since, as discussed in \S\ref{sec:theory} the time
derivative is close to zero for most of the evolution of the Universe
and only weakly evolving away from this at later times. On smaller
scales this may be a less reasonable approximation, however, we are
still using the fully nonlinear gravitational potential field.

%%%%%%%%%%%%%%%%%%%%%%%%%%%%%%%%%%%%%%%%%%%%%%%%%%%%%%%

\end{document}